\documentclass[conference]{IEEEtran}
\IEEEoverridecommandlockouts

\usepackage{amsfonts}
\usepackage{amsthm}
\usepackage{amsmath}
\usepackage{bbm}
\usepackage{amssymb}
\usepackage{algorithm}
\usepackage{algpseudocode}
\usepackage{enumerate}
\usepackage{graphicx}
\usepackage{booktabs}
\usepackage{footnote}
\usepackage{multirow}
\usepackage{flushend}
\usepackage{extarrows}
\usepackage{xcolor}
\usepackage{subfigure}
\usepackage{makecell}
\usepackage{cite}
\usepackage{enumitem}
\usepackage{fancyhdr}

\usepackage[breaklinks=true]{hyperref}

\newtheorem{theorem}{Theorem}
\newtheorem{lemma}{Lemma}
\newtheorem{definition}{Definition}

\def\BibTeX{{\rm B\kern-.05em{\sc i\kern-.025em b}\kern-.08em
		T\kern-.1667em\lower.7ex\hbox{E}\kern-.125emX}}

\begin{document}
%
\title{LDPRecover: Recovering Frequencies from Poisoning Attacks against Local Differential Privacy}

\author{
	\IEEEauthorblockN{
		Xinyue Sun$^{1,2}$,  Qingqing Ye$^2$, Haibo Hu$^2$, Jiawei Duan$^2$, Tianyu Wo$^1$, Jie Xu$^3$, Renyu Yang$^1$\thanks{Qingqing Ye and Tianyu Wo are the corresponding authors.}
	}
	\IEEEauthorblockA{
		$^1$Beihang University; $^2$The Hong Kong Polytechnic University; $^3$The University of Leeds
	}
	\IEEEauthorblockA{
		\{xy.sun, woty, renyuyang\}@buaa.edu.cn; \{qqing.ye, haibo.hu\}@polyu.edu.hk, jiawei.duan@connect.polyu.hk; j.xu@leeds.ac.uk
	}
}

\maketitle
\thispagestyle{fancy}
\fancyhead[C]{\textcolor{red}{This paper has been accepted by IEEE 40th Annual International Conference on Data Engineering (ICDE2024)}}

\begin{abstract}
Local differential privacy (LDP), which enables an untrusted server to collect aggregated statistics from distributed users while protecting the privacy of those users, has been widely deployed in practice. However, LDP protocols for frequency estimation are vulnerable to poisoning attacks, in which an attacker can poison the aggregated frequencies by manipulating the data sent from malicious users. Therefore, it is an open challenge to recover the accurate aggregated frequencies from poisoned ones.

In this work, we propose \textit{LDPRecover}, a method that can recover accurate aggregated frequencies from poisoning attacks, even if the server does not learn the details of the attacks. In LDPRecover, we establish a genuine frequency estimator that theoretically guides the server to recover the frequencies aggregated from genuine users' data by eliminating the impact of malicious users' data in poisoned frequencies. Since the server has no idea of the attacks, we propose an adaptive attack to unify existing attacks and learn the statistics of the malicious data within this adaptive attack by exploiting the properties of LDP protocols. By taking the estimator and the learning statistics as constraints, we formulate the problem of recovering aggregated frequencies to approach the genuine ones as a \textit{constraint inference} (CI) problem. Consequently, the server can obtain accurate aggregated frequencies by solving this problem optimally. Moreover, LDPRecover can serve as a frequency recovery paradigm that recovers more accurate aggregated frequencies by integrating attack details as new constraints in the CI problem. Our evaluation on two real-world datasets, three LDP protocols, and untargeted and targeted poisoning attacks shows that LDPRecover is both accurate and widely applicable against various poisoning attacks.
\end{abstract}
 \section{Introduction}
Local differential privacy (LDP)~\cite{duchi2013local}, a variant of differential privacy~\cite{dwork2006calibrating,dwork2014algorithmic}, is an emerging paradigm that enables an untrusted server to gather aggregated statistics from distributed users while providing provable privacy protection for these users. In LDP, participating users perturb their data locally and report the perturbed data to the untrusted server. Then the server aggregates the statistics of interest from these perturbed data. Thanks to its rigorous privacy guarantee, LDP has been widely deployed in practice. For example, Google~\cite{erlingsson2014rappor,fanti2015building,bittau2017prochlo} has integrated LDP in Chrome to collect default homepages and search engines; Apple~\cite{appledf2017} gathers popular emojis and words by deploying LDP in IOS.

\begin{figure}[tb]
	\setlength{\abovecaptionskip}{-0.1em}
	\setlength{\belowcaptionskip}{-0.1em}
	\centerline{\includegraphics[width=\linewidth]{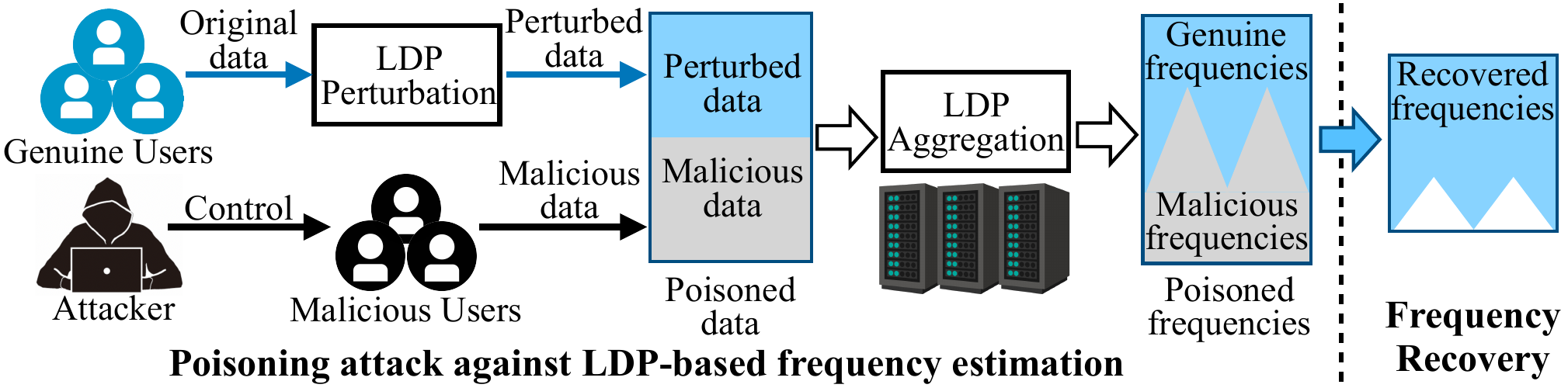}}
	\caption{Illustration of poisoning attack against LDP-based frequency estimation and our frequency recovery.}
	\label{fig_poisoning_attacks}
\end{figure}

However, due to its distributed settings, LDP is vulnerable to \textit{poisoning attacks}~\cite{cheu2021manipulation,cao2021data,wu2022poisoning}, where an attacker may hijack users or inject malicious users to corrupt the LDP protocol.
For example, the attacker poisons the aggregated frequencies of arbitrary items by crafting the data sent to the server from these malicious users, as shown in Figure \ref{fig_poisoning_attacks}.
Poisoning attacks against LDP protocols can be further divided into \textit{untargeted}~\cite{cheu2021manipulation} and \textit{targeted poisoning attacks}~\cite{cao2021data,wu2022poisoning}. In the untargeted attacks, the attacker aims to degrade the overall accuracy of the aggregated frequencies for all items. In the targeted attacks, on the other hand, the attacker wants to increase the aggregated frequencies of attacker-chosen items (i.e., target items) and thus promote them as popular items. Regardless of poisoning types, the server needs to recover accurate aggregated frequencies from the poisoned ones.

In the literature, frequency recovery in the LDP protocol suffering from poisoning attacks is largely unexplored. Although there are a few countermeasures~\cite{cao2021data} against targeted attacks, such as detecting malicious users, they are only effective on specific attacks (e.g., MGA~\cite{cao2021data}) and require the server to know the details of these attacks. 

\vspace{0.3em} \noindent \textbf{Our Contributions.} In this work, we propose \textit{LDPRecover}, a method that can recover accurate aggregated frequencies from the poisoned ones under LDP protocols, even if the details of the attacks are unknown to the server. While LDPRecover works against both untargeted and targeted attacks by enhancing the overall accuracy of items' aggregated frequencies, for the targeted attack (e.g., MGA~\cite{cao2021data}), LDPRecover can also reduce the frequency gains (i.e., the increase in frequency) by the attacker on the target items. 

LDPRecover is built on our insight that in a poisoning attack, the \textit{poisoned frequencies} aggregated by the server are mixture of \textit{genuine frequencies} aggregated from genuine users' data and \textit{malicious frequencies} aggregated from malicious users' data, as shown in Figure \ref{fig_poisoning_attacks}. As such, the server can recover genuine frequencies by deducting malicious frequencies from the poisoned frequencies. However, this idea poses non-trivial challenges. First, the theoretical relationship between the distributions of poisoned, genuine, and malicious frequencies is unexplored. Second, even if this relationship could be derived, the server has yet no prior information about malicious users to recover the genuine frequencies.

To address these challenges, we first propose an analytical framework that generalizes the poisoning attacks against LDP protocols, from which we derive the theoretical relationship between poisoned, genuine, and malicious frequencies. 
On this basis, we establish a genuine frequency estimator that guides the server to recover the genuine frequencies from the poisoned ones. 
Then, we propose an adaptive attack to unify state-of-the-art untargeted and targeted attacks, in which we can  learn the statistics of malicious frequencies by leveraging the aggregated properties of LDP protocols. 
By taking both the genuine frequency estimator and the learnt malicious statistics as constraints, we formulate the problem of recovering aggregated frequencies to approach the genuine ones as an \textit{constraint inference (CI) problem}, whose objective function is to minimize the $L_2$ norm between the recovered and genuine frequencies. Thus, LDPRecover can recover the aggregated frequencies with high accuracy by solving the CI problem optimally.

While LDPRecover does not depend on any specific details about the attacks, these details can help LDPRecover to recover more accurate aggregated frequencies by integrating them as new constraints in the CI problem. 
For example, when a targeted attack causes a significant increase in target items' frequencies, conventional outlier detection techniques~\cite{kieu2018outlier,zhou2018non,su2019robust} can identify these target items by detecting statistical anomalies in the historical frequency data of each item.
In this case, LDPRecover can exploit such knowledge of target items to recover more accurate aggregated frequencies.

We empirically evaluate our proposed LDPRecover using two datasets, three popular LDP protocols (i.e., GRR~\cite{kairouz2014extremal}, OUE, and OLH~\cite{wang2017locally}), as well as three poisoning attacks to LDP (i.e., an untargeted poisoning attack called Manip~\cite{cheu2021manipulation}, a targeted poisoning attack called MGA~\cite{cao2021data}, and our proposed adaptive attack called AA). Results show that LDPRecover not only recovers accurate aggregated frequencies from the poisoned ones but also substantially reduces the frequency gains of the targeted poisoning attacks.
Our contributions can be summarized as follows:
\begin{itemize}
	\item To the best of our knowledge, we are the first to put forward a systematic study on frequency recovery from LDP protocols that suffer from general poisoning attacks.
	\item We propose an LDPRecover method to recover the accurate aggregated frequencies from poisoned ones, even if the server does not learn the details of the attacks. Furthermore, LDPRecover can serve as a frequency recovery paradigm that further improves the accuracy of the recovered frequencies by integrating attack details.
	\item We evaluate the effectiveness of LDPRecover for three popular LDP protocols suffering from poisoning attacks on two real-world datasets. Results reveal that our proposed method can effectively recover accurate aggregated frequencies and counter these poisoning attacks.
\end{itemize}

The rest of the paper is organized as follows. We discuss related work in Section \ref{Sec_related}, provide the preliminaries in Section \ref{Sec_preliminaries}, and formulate our problem statement in Section \ref{Sec_problem_definition}. Then, we  present our recovery method called LDPRecover in Section \ref{Sec_LDPRecover} and show experimental results in Section \ref{Sec_evaluate}. Finally, we further discuss the applicability of LDPRecover in Section \ref{Sec_discussion} and conclude this paper in Section \ref{Sec_conclusion}.

\section{Related work}\label{Sec_related}
\noindent \textbf{Local Differential Privacy.} Local differential privacy (LDP)~\cite{dwork2006calibrating,duchi2013local,dwork2014algorithmic} has become a \textit{de facto} standard privacy model in sensitive data collection and analysis. In LDP, the server, which aims to collect private data from users, is considered untrusted. Each user first locally perturbs her data by a certain LDP mechanism (e.g., GRR~\cite{kairouz2014extremal}) before sending it to an untrusted server. Based on the perturbed data from users, the server can derive certain aggregated statistics without seeing the actual private data of each user. Due to its rigorous privacy properties, LDP has been widely studied in various tasks, including frequency estimation~\cite{warner1965randomized,kairouz2014extremal,wang2017locally,bassily2015local,kairouz2016discrete,zhang2018calm,jia2019calibrate,wang2019consistent,xue2022ddrm}, mean estimation~\cite{ding2017collecting,duchi2018minimax,wang2019collecting,li2020estimating,duan2022utility}, heavy hitters identification~\cite{bassily2017practical,wang2019locally,qin2016heavy,wang2018locally}, range queries~\cite{li2014data,cormode2019answering, yang2020answering}, and other more complex tasks~\cite{he2015dpt,chen2016private,qin2017generating,avent2017blender,cormode2018marginal,ren2018textsf,gursoy2018utility,wang2019answering,ye2020towards,ye2020lf,cunningham2021real}. 

\vspace{0.3em} \noindent \textbf{Poisoning Attack against LDP.} LDP is vulnerable to poisoning attacks~\cite{cheu2021manipulation,cao2021data,wu2022poisoning}, in which an attacker poisons the server's aggregated statistics by manipulating the data sent from malicious users. Depending on the attackers' goal, poisoning attacks against LDP protocols be categorized into untargeted poisoning attacks~\cite{cheu2021manipulation} and targeted poisoning attacks~\cite{cao2021data,wu2022poisoning}. In this work, we focus on countering poisoning attacks against LDP protocols for frequency estimation, as these protocols can serve as the building block of more advanced tasks. Thus, we review Manip~\cite{cheu2021manipulation}, which is a popular untargeted poisoning attack, and MGA~\cite{cao2021data}, which is a popular targeted poisoning attack. Specifically, Manip seeks to distort the distribution of aggregated frequencies in $L_1$ norm. Conversely, MGA strives to amplify the frequency gains of items chosen by the attacker, where ``frequency gain'' denotes the increase in the target item's frequency after the targeted attack. Note that in both attacks, the attacker requires that malicious users directly send the attacker-crafted data to the server, as this is a more effective way in terms of attack results~\cite{cheu2021manipulation, cao2021data}.

\vspace{0.3em} \noindent \textbf{Countermeasures against Poisoning Attacks to LDP.} We note that several countermeasures were proposed to counter the targeted poisoning attack (i.e., MGA), including malicious users detection and conditional probability based detection \cite{cao2021data}. Specifically, these countermeasures, which only apply to OUE and OLH, are built on strong assumptions, e.g., the server knows the details of this attack. However, in reality, these assumptions do not always hold, resulting in these countermeasures being invalid in most cases. As for the untargeted poisoning attack, there is no study to deal with it yet.

\section{Preliminaries}\label{Sec_preliminaries}
\subsection{Local Differential Privacy}
In LDP~\cite{duchi2013local}, there are many users and one server. Each user possesses an item (data) $v$ from a domain $D$, and the server, which is not trusted by users, wants to learn statistics among all users' data. To protect privacy, each user perturbs her input item $v \in D$ locally with an algorithm $\Psi(\cdot)$ and sends the perturbed data $\Psi(v)$ to the server. Formally, the privacy requirement is that $\Psi(\cdot)$ satisfies the following property.
\begin{definition}[$\epsilon$-Local Differential Privacy \cite{duchi2013local}]
	A randomized algorithm $\Psi(\cdot)$ satisfies $\epsilon$-LDP, where $\epsilon >=0$, if and only if for any two input $v_1, v_2 \in D$, we have 
	\begin{equation}
		\forall T \subseteq Range(\Psi): \ \Pr[\Psi(v_1) \in T] \leq e^{\epsilon}\Pr[\Psi(v_2) \in T], 
	\end{equation}
	where $Range(\Psi)$ denotes the set of all possible outputs of $\Psi$. 
\end{definition}
The offered privacy is controlled by privacy budget $\epsilon$, i.e., a smaller (resp. larger) $\epsilon$ implies a stronger (resp. weaker) privacy level.

\subsection{LDP Protocols for Frequency Estimation}\label{sec_LDP_protocols}
We review three state-of-the-art pure LDP protocols for frequency estimation. These protocols can be specified by a pair of algorithms $(\Psi, \Phi)$: each user uses $\Psi$ to perturb her input item, and the server uses $\Phi$ to aggregate the items' frequencies in the perturbed data sent from the users.

\vspace{0.3em} \noindent \textbf{General Randomized Response (GRR).} General Randomized Response (GRR)~\cite{kairouz2014extremal}, a generalized version of randomized response~\cite{warner1965randomized}, is a basic protocol in LDP. In GRR, each user sends her true item $v \in D$ to the untrusted server with probability $p$ or sends a random $v' \neq v$ with probability $q$. Formally, the perturbation algorithm $\Psi_{\text{GRR}(\epsilon)}$ is defined as
\begin{equation}\label{equ_grr}
	\Pr[\Psi_{\text{GRR}(\epsilon)}(v)= b] =\begin{cases} \frac{e^{\epsilon}}{d-1+e^{\epsilon}} \triangleq p ,&  \text{if } b = v, \\ \frac{1}{d-1+e^{\epsilon}} \triangleq q,&  \text{if } b \neq v, \end{cases}
\end{equation}
where $d$ is the size of $D$, i.e., $d=|D|$. It is easy to prove this satisfies $\epsilon$-LDP since $\frac{p}{q}=e^{\epsilon}$. To estimate the frequency of $v \in D$, the server first counts $v$, denoted by $C(v)$, then computes the estimated count of the users who have $v$ as private item:
\begin{equation}
	\Phi_{\text{GRR}(\epsilon)}(v) := \frac{C(v)- nq}{p-q}, 
\end{equation}
where $n$ is the total number of users. Then the estimated frequency of $v$ is $\tilde{f}(v) = \frac{1}{n}\Phi_{\text{GRR}(\epsilon)}(v)$. In \cite{wang2017locally}, it is shown that $\Phi_{\text{GRR}(\epsilon)}(\cdot)$ is an unbiased estimation of true counts, and the variance of this estimation is 
\begin{equation}
	\mathbf{Var}[\Phi_{\text{GRR}(\epsilon)}(v)] = n \cdot \frac{d-2 +e^{\epsilon}}{(e^{\epsilon}-1)^2} + n f(v) \cdot \frac{d-2}{e^{\epsilon}-1}
\end{equation}

\vspace{0.3em} \noindent \textbf{Optimized Unary Encoding (OUE).} Optimized Unary Encoding (OUE) protocol is designed to avoid the variance of the estimation depending on the domain size $d$ by encoding the item into the unary representation. In OUE, each user first encodes her item $v \in D$ to a $d$-bit binary vector $\mathbf{b}=[b_1 \ b_2 \ \cdots \ b_d]$ whose bits are all $0$ except for the $1$ in the $v$-th bit. Then, each user perturbs each bit of her encoded binary vector with $\Phi_{\text{OUE}(\epsilon)}(\cdot)$ independently. Specifically, 
\begin{equation}
	\Pr[\Psi_{\text{OUE}(\epsilon)}(b_i) = 1] = 
	\begin{cases} 
		\frac{1}{2} \triangleq p, & \text{if } i=v, \\
		\frac{1}{e^{\epsilon}+1} \triangleq q, & \text{otherwise},
	\end{cases}
\end{equation}
where $\tilde{b}_i = \Psi_{\text{OUE}(\epsilon)}(b_i)$ is the $i$-th perturbed bit, and $\mathbf{\tilde{b}} = [\tilde{b}_1 \ \tilde{b}_2 \ \cdots \ \tilde{b}_d]$ is the perturbed vector.

Given the reports $\mathbf{\tilde{b}}^j$ from all users $j \in [n]$, to estimate the frequency of $v$, the server counts the number of reports whose $v$-th bit is set to $1$, denoted by $C(v)=|\{j|\tilde{b}_v^j =1\}|$. Then, the server transforms $C(v)$ to its unbiased estimation by 
\begin{equation}
	\Phi_{\text{OUE}(\epsilon)}(v) := \frac{C(v) - nq}{p-q}.
\end{equation}
It is proven in \cite{wang2017locally} that $\Phi_{\text{OUE}(\epsilon)}(\cdot)$ satisfies $\epsilon$-LDP, and the estimated count is unbiased and has variance
\begin{equation}
	\mathbf{Var}[\Phi_{\text{OUE}(\epsilon)}(v)] = n \cdot \frac{4e^{\epsilon}}{(e^{\epsilon}-1)^2}
\end{equation}

\vspace{0.3em} \noindent \textbf{Optimized Local Hashing (OLH).} Optimized Local Hashing (OLH) \cite{wang2017locally, wang2019consistent} protocol aims to deal with a large domain size $d$ by applying a hash function to map an input item into a smaller domain of size $g$ (i.e., $g \ll d$). In particular, OLH sets $g$ to $\lceil e^{\epsilon}+1 \rceil$ by default, as it achieves the lowest variance with this setting. In OLH, each user first randomly picks a hash function $H$ from a family of hash functions $\mathbf{H}$ (e.g., xxhash) and then computes the hashed value of her item $v$ as $b = H(v)$, where $b \in \{1,2,...,g\}$ is a value hashed from $v \in D$ using $H$, and the tuple $(H, b)$ is the encoded value for $v$. Note that $\mathbf{H}$ should have the property that the distribution of each $v$'s hash value is uniform over $\{1,2,...,g\}$ and independent from the distributions of other input items in $D$. Next, each user perturbs her encoded value by the following perturbation function.
\begin{equation}
	\Psi_{\text{OLH}(\epsilon)}(v):= \left\langle H, \Psi_{\text{GRR}(\epsilon)}(H(v)) \right\rangle.
\end{equation}
where $\Psi_{\text{GRR}(\epsilon)}(\cdot)$ is the perturbation algorithm of GRR on the domain $\{1,2,...,g\}$.

Let $\left\langle H^j, b^j \right\rangle$ be the report from the $j$-th user. To estimate the frequency of $v \in D$, the server first counts the number of reports whose input item could be $v$, denoted by $C(v) = \{j|H^j(x)=a^j\}$. Then, the server transforms $C(v)$ to its unbiased estimation
\begin{equation}
	\Phi_{\text{OLH}(\epsilon)}(v) := \frac{C(v)- nq}{p-q},
\end{equation}
where $p=e^{\epsilon}/(e^{\epsilon}+g-1)$ and $q =1/g$. The variance of this estimation is 
\begin{equation}
	\mathbf{Var}[\Phi_{\text{OLH}(\epsilon)}(v)] = n \cdot \frac{4e^{\epsilon}}{(e^{\epsilon}-1)^2}.
\end{equation}

\subsection{Summary of Common Properties of LDP Protocols}\label{sec_properties_LDP}
Here we summarize the common properties of the pure LDP protocols. 
When the server aggregates the reports from all users, for each item $v \in D$, its estimated count for any LDP protocol can be represented in a unified way:
\begin{equation}\label{aggregation_fuction}
	\Phi_{\epsilon}(v) := \frac{C(v)- nq}{p-q}.
\end{equation}
Note that the perturbed probabilities $p$ and $q$ in various protocols are different.
Besides, since these protocols are pure LDP protocols~\cite{wang2017locally}, $C(v)$ can be represented as follows.
\begin{equation}
	C(v)=\sum\nolimits_{i=1}^n \mathbbm{1}_{S(\tilde{v}_i)}(v)
\end{equation}
where $\mathbbm{1}_{S(\tilde{v}_i)}(v)$ is a characteristic function:
\begin{equation}\label{equ_characteristic_function}
	\mathbbm{1}_{S(\tilde{v}_i)}(v) =\begin{cases}
		1, & \text{ if } v \in S(\tilde{v}_i),\\
		0, & \text{otherwise.}
	\end{cases}
\end{equation}
In particular, $\tilde{v}_i$ denotes the perturbed data of the $i$-th user, and $S(\tilde{v}_i)$ denotes the set of items that $\tilde{v}_i$ supports, i.e., the support $S(\tilde{v}_i)$ of a perturbed item $\tilde{v}_i$ is the set of items whose encoded values could be $\tilde{v}_i$.

\section{Problem Definition}\label{Sec_problem_definition}
\subsection{Threat Model}
We focus on the threat model in prior studies of poisoning attacks against LDP protocols~\cite{cheu2021manipulation,cao2021data}. In what follows, we discuss the attacker's goals, capabilities, and background knowledge in detail.

\vspace{0.3em} \noindent \textbf{Attacker's goals.} In an untargeted poisoning attack, the attacker's goal is to indiscriminately increase the error in the frequencies of items aggregated by the server. In a targeted poisoning attack, the attacker aims to increase the frequencies of the target items chosen by the attacker. 

\vspace{0.3em} \noindent \textbf{Attacker's Capabilities and Background Knowledge.} We assume an attacker can control some malicious users in an LDP protocol. These malicious users could be fake users injected into the LDP protocol or genuine users compromised by the attacker. The attacker crafts the malicious data sent from these malicious users to the server.

Since the LDP protocol is executed on the users' side, the attacker knows the details of the LDP protocol adopted by the genuine users. Specifically, the attacker knows various parameters of the LDP protocols, including input domain $D$, encoded domain $\tilde{D}$, and privacy budget $\epsilon$. 

\subsection{Design Goals}
We aim to design an accurate and widely-applicable frequency recovery method for LDP protocols that suffer from poisoning attacks. Even without knowledge about the poisoning attacks, our recovery method should still be able to recover aggregated frequencies close to the genuine ones. Specifically, our design goals are as follows.

\vspace{0.3em} \noindent \textbf{Accuracy.} The aggregated frequencies recovered by our recovery method should be accurate. For both untargeted and targeted poisoning attacks, the recovered frequencies should be close to the genuine frequencies that the server gathers from genuine users using an LDP protocol. Furthermore, for the targeted poisoning attack, we require that the frequency gains of that attack in the aggregated frequencies recovered by our method should be very low.

\vspace{0.3em} \noindent \textbf{Applicability.} Our recovery method should be widely applicable to counter the poisoning attacks to LDP protocols, even if the server does not know the details of the attacks. In particular, the server may acquire partial knowledge about the attacks in some applications. 
For example, the server, utilizing outlier detection methodologies~\cite{kieu2018outlier,zhou2018non,su2019robust}, can deduce the attacker's target items through careful analysis of historical data.
In this case, our recovery method should be able to recover more accurate frequencies by exploiting such knowledge of these target items.

\section{LDPRecover}\label{Sec_LDPRecover}
\subsection{Overview}\label{sec_overview}
LDPRecover is based on the following insights. Note that in a poisoning attack, the poisoned frequencies are mixture of genuine and malicious frequencies. Suppose we have a genuine frequency estimator that enables the server to recover the genuine frequencies by deducting the malicious frequencies from the poisoned ones. Then, if we can learn the malicious frequencies of items, the server can recover the aggregated frequencies close to the genuine ones from the poisoned ones. Following the insights, we design LDPRecover with three major parts: estimator construction, malicious frequency learning, and genuine frequency recovery. 

\vspace{0.3em} \noindent \textbf{Step 1: Estimator Construction.} The first step of LDPRecover is to construct a genuine frequency estimator, which guides the server on how to recover the genuine frequencies. As such, we first propose an analytical framework to generalize the poisoning attacks against LDP protocols, by which we further derive the theoretical relationship between poisoned, genuine, and malicious frequencies. On this basis, we establish the genuine frequency estimator and analyze the expectation and variance of the estimator. We provide the details of this step in Section \ref{sec_esitmator} and further give the error analysis of this estimator in Section \ref{sec_error_analysis}.

\vspace{0.3em} \noindent \textbf{Step 2: Malicious Frequency Learning.} Since we assume that the server has no details of the poisoning attacks, we cannot obtain the malicious frequencies of items directly. To tackle this challenge, we propose an adaptive attack that unifies state-of-the-art untargeted and targeted attacks~\cite{cheu2021manipulation,cao2021data}. 
Following this, we leverage the aggregated properties of LDP protocols to learn the statistics of malicious frequencies, specifically their summation, within the adaptive attack. These learning statistics serve as an alternative approximation for the malicious frequencies. We give the details of this part in Section \ref{sec_malicious_frequecny_learning}.

\vspace{0.3em} \noindent \textbf{Step 3: Genuine Frequency Recovery.} By treating the genuine frequency estimator and the learning statistics of malicious frequencies as constraints, we formulate the problem of recovering aggregated frequencies to approach the genuine ones as a Constraint Inference (CI) problem. The server can obtain accurate aggregated frequencies by solving this CI problem. Note that both the genuine frequency estimator and the statistics of the malicious frequencies are derived from the public information known to the server, such as the LDP protocols. Therefore, LDPRecover can work even if the server has no details of the attacks. More importantly, when the server is able to acquire partial knowledge about the attacks, LDPRecover can also exploit such knowledge to help the server better counter the attacks and improve the accuracy of the recovered frequencies. More details of this step are illustrated in Section \ref{sec_genuine_frequency_recovery}.

\subsection{Estimator Construction}\label{sec_esitmator}
This subsection corresponds to Step 1 of LDPRecover. Here we present a general framework for poisoning attacks against LDP protocols, by which we further establish the genuine frequency estimator. 

\subsubsection{Analytical Framework for Poisoning Attacks}\label{sec_framework}
\begin{figure}[tb]
	\centerline{\includegraphics[width=\linewidth]{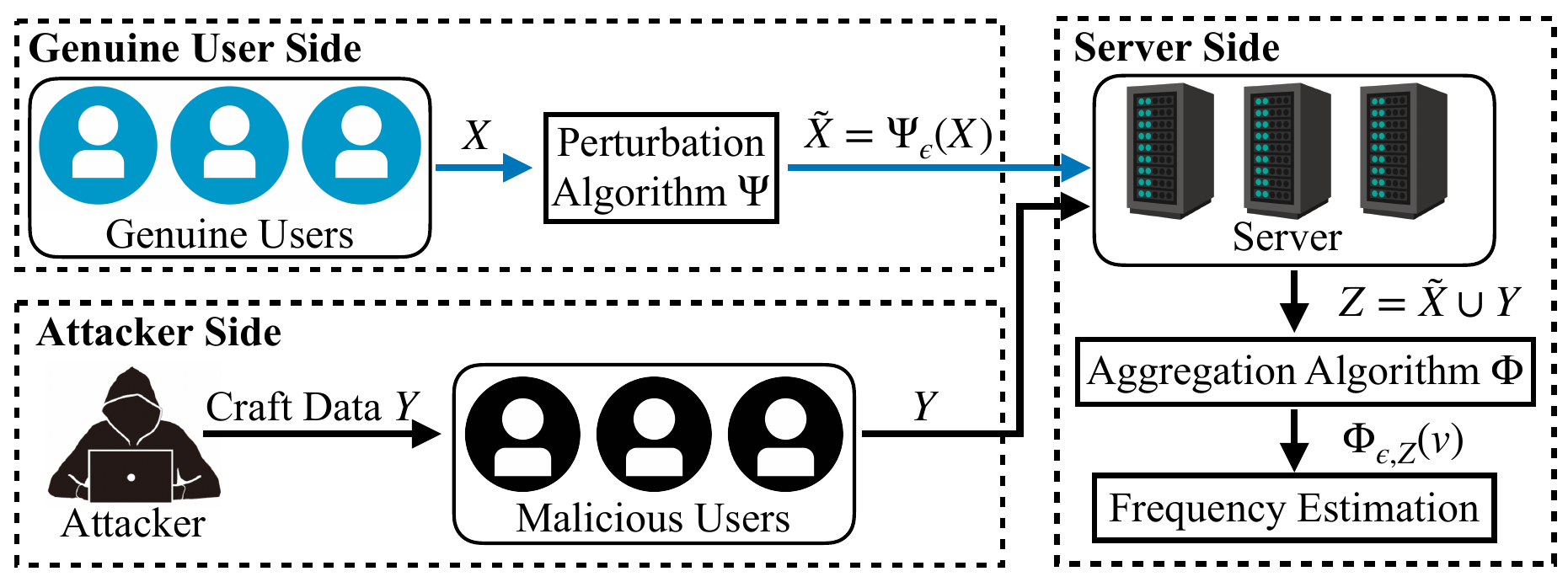}}
	\caption{The general process of poisoning attacks against LDP protocols.}
	\label{fig_framework_overview}
\end{figure} 
Our framework includes three parties: genuine users, an attacker, and a server. Figure~\ref{fig_framework_overview} summarizes our framework. 
\begin{itemize}[leftmargin=*]
	\item \textbf{Genuine User Side:} Suppose there are $n$ genuine users. Each user possesses a private item $x$ and perturbs it into $\tilde{x}=\Psi_{\epsilon}(x)$ using the perturbation algorithm $\Psi_{\epsilon}(\cdot)$ of an LDP protocol with a privacy budget $\epsilon$. Then, the genuine users report the perturbed data to the server. In particular, $\tilde{x}$ could be an index (e.g., for GRR), a binary vector (e.g., for OUE), and a tuple (e.g., for OLH). We use $x_i \in D$ and $ \tilde{x}_i \in \tilde{D}$ to denote the original item and perturbed data of the $i$-th user, where $D$ and $\tilde{D}$ denote the input domain and encoded domain of the LDP protocol, respectively. Moreover, we use $X=\{x_1, x_2, ..., x_n\}$ and $\tilde{X}=\{\tilde{x}_1, \tilde{x}_2, ..., \tilde{x}_n\}$ to denote the original items and perturbed data of $n$ genuine users, respectively.  
	
	\item \textbf{Attacker Side:} The attacker crafts data in $\tilde{D}$ for $m$ malicious users, and these users directly send the attacker-crafted data to the server. We use $Y=\{y_1, y_2,..., y_m\}$ to denote the attacker-crafted data sent from these malicious users.
	
	\item \textbf{Server Side:} Once receiving the reported data from genuine and malicious users, the server could estimate the count of each item $v \in D$ from the reported data. Let $Z=\{z_1 ,z_2 ,..., z_{n+m}\}$ denote the reported data, i.e., $Z = \tilde{X} \cup Y$, and $\Phi_{\epsilon, Z}(v)$ be the estimated count of $v$ in $Z$, where $\Phi_{\epsilon}(\cdot)$ denotes the aggregation algorithm executed on $Z$.
\end{itemize}

Under this framework, we can analyze the theoretical relationship between the frequency of each item in $\tilde{X}$, $Y$, and $Z$. Specifically, for each item $v \in D$, we use $\tilde{f}_{\tilde{X}}(v)$, $\tilde{f}_{Y}(v)$ and $\tilde{f}_Z(v)$ to denote the \textit{genuine}, \textit{malicious}, and \textit{poisoned frequency} of $v$, where $\tilde{f}_{\tilde{X}}(v)$, $\tilde{f}_{Y}(v)$ and $\tilde{f}_Z(v)$ are \textbf{aggregated} from $\tilde{X}$, $Y$, and $Z$, respectively, using an LDP protocol. The relationship between $\tilde{f}_{\tilde{X}}(v)$, $\tilde{f}_{Y}(v)$ and $\tilde{f}_Z(v)$ can be represented as follows. 

\begin{equation}
	\tilde{f}_{Z}(v) = \frac{n}{n+m} \tilde{f}_{\tilde{X}}(v) + \frac{m}{n+m} \tilde{f}_{Y}(v). \label{equ_relationship_xyz}
\end{equation}

This equation implies that $\tilde{f}_Z(v)$ can be decomposed into a linear combination of $\tilde{f}_{\tilde{X}}(v)$ and $\tilde{f}_{Y}(v)$. However, due to the randomness of LDP protocols, $\tilde{f}_{\tilde{X}}(v)$ and $\tilde{f}_{Y}(v)$ are distributions but not deterministic values. Therefore, we need to establish the relationship of the distributions of $\tilde{f}_{\tilde{X}}(v)$, $\tilde{f}_{Y}(v)$, and $\tilde{f}_Z(v)$. To be specific, for the data sent from genuine users and malicious users (i.e., $\tilde{X}$ and $Y$), we use $\Phi_{\epsilon, \tilde{X}}(v)$  and $\Phi_{\epsilon, Y}(v)$ to denote the estimated counts of $v$ in $\tilde{X}$ and $Y$, respectively. Consequently, $\tilde{f}_{\tilde{X}}(v)$ and $\tilde{f}_{Y}(v)$ can be expressed as:
\begin{equation}
	\tilde{f}_{\tilde{X}}(v) = \frac{1}{n} \Phi_{\epsilon, \tilde{X}}(v), \ 
	\tilde{f}_{Y}(v) = \frac{1}{m} \Phi_{\epsilon, Y}(v).
\end{equation}

Recalling the properties of LDP protocols mentioned in Section \ref{sec_properties_LDP}, both $\Phi_{\epsilon, \tilde{X}}(v)$ and $\Phi_{\epsilon, Y}(v)$ can be decomposed as follows. 
\begin{equation}
	\Phi_{\epsilon, \tilde{X}}(v) = \sum_{\tilde{x} \in \tilde{X}}  \frac{\mathbbm{1}_{S(\tilde{x})}(v)- q }{p-q},  \ 
	\Phi_{\epsilon, Y}(v) =\sum_{y \in Y}   \frac{\mathbbm{1}_{S(y)}(v)- q }{p-q},
\end{equation}
where $p$ and $q$ represent the perturbation probabilities of the LDP protocol. Without loss of generality, we slightly abuse $\Phi_{\epsilon, \tilde{x}}(v)$ and $\Phi_{\epsilon, y}(v)$ to denote the estimated count of $v$ in any data $\tilde{x}$ and $y$, respectively, as follows:
\begin{align}\label{equ_agg_xy}
	\Phi_{\epsilon, \tilde{x}}(v)=\frac{\mathbbm{1}_{S(\tilde{x})}(v)- q }{p-q}, \ \ \ \Phi_{\epsilon, y}(v)=\frac{\mathbbm{1}_{S(y)}(v)- q }{p-q},
\end{align}
where $\mathbbm{1}_{S(\cdot)}(v)$ is the characteristic function defined in Equation (\ref{equ_characteristic_function}). Hence, $\Phi_{\epsilon, \tilde{X}}(v)$ and $\Phi_{\epsilon, Y}(v)$ can be redefined as:
\begin{align}\label{equ_sum_agg_xy}
	\Phi_{\epsilon, \tilde{X}}(v) \! = \! \sum\nolimits_{\tilde{x} \in \tilde{X}} \Phi_{\epsilon, \tilde{x}}(v), \Phi_{\epsilon, Y}(v) = \sum\nolimits_{y \in Y}\!\Phi_{\epsilon, y}(v).
\end{align}
Building on this groundwork, we can model the distributions of $\tilde{f}_{\tilde{X}}(v)$ and $\tilde{f}_{Y}(v)$ using \textit{Lindeberg–L\'evy Central Limit Theorem} (CLT)~\cite{shanthikumar1984central,fischer2011history}.

\vspace{0.3em} \noindent \textbf{Model the Distribution of $\tilde{f}_{Y}(v)$.} We first model the distribution of $\tilde{f}_{Y}(v)$ in the poisoning attack. Note that $\Phi_{\epsilon, Y}(v)$ could be regarded as the summation of $m$ independent identically distributed random variables (i.e., $\Phi_{\epsilon, y}(v)$) \footnote{The process of an attacker crafting data for a malicious user is essentially equivalent to sampling from the distribution specified by the attacker (See Section \ref{sec_malicious_frequecny_learning}). Thus, the aggregated results of each sample (crafted data) still follow the same distribution, as all samples are executed by the same aggregation algorithm.}. Therefore, the following lemma establishes the asymptotic distribution of $\tilde{f}_{Y}(v)$.

\begin{lemma}\label{lmm_y_normal}
	The asymptotic distribution $\tilde{f}_{Y}(v)$ is $\mathcal{N}(\mu_y, \sigma_y^2)$, $\lim\limits_{m \to \infty} \tilde{f}_{Y}(v) \sim \mathcal{N}(\mu_y, \sigma_y^2)$, where $\mathcal{N}$ denotes a normal distribution, $\mu_y = \mathbf{E}[\Phi_{\epsilon, y}(v)]$, and $\sigma_y^2= \mathbf{Var}[\Phi_{\epsilon, y}(v)] / m $.
\end{lemma}
\begin{proof}
	Please refer to our technical report \cite{sunxy}.
\end{proof}

\vspace{0.3em} \noindent \textbf{Model the Distribution of $\tilde{f}_{\tilde{X}}(v)$.} It is rather challenging to model the distribution of $\tilde{f}_{\tilde{X}}(v)$. Indeed, $\tilde{f}_{\tilde{X}}(v)$ is the estimated frequency of $v$ using the complete algorithm pair (i.e., $\Psi$ and $\Phi$) of the LDP protocol on $X$. However, each genuine user perturbs her original data $x$ to $\tilde{x} = \Psi_{\epsilon}(x)$, and different original data follow different perturbations (see Section~\ref{sec_LDP_protocols}). Thus, $\{\Phi_{\epsilon, \tilde{x}}(v)| \tilde{x} \in \tilde{X} \}$ are probably not identically distributed, which does not satisfy the prerequisite of \textit{Lindeberg–L\'evy} CLT. 

Fortunately, it is still viable to model $\tilde{f}_{\tilde{X}}(v)$ as one normal distribution. This is because genuine users with identical input data apply the same perturbation algorithm, leading to independently and identically distributed estimated counts among these users. As such, the perturbed data arising from identical input data can be partitioned into distinct subsets. Following this, one normal distribution can be deployed to approximate the summation of $v$'s aggregated frequencies in each subset. The asymptotic distribution of \(\tilde{f}_{\tilde{X}}(v)\) is formalized in the following lemma.

\begin{lemma}\label{lmm_x_normal}
	The asymptotic distribution of $\tilde{f}_{\tilde{X}}(v)$ is $\mathcal{N}(\mu_{\tilde{x}}, \sigma_{\tilde{x}}^2)$, i.e., $\lim\limits_{n \to \infty} \tilde{f}_{\tilde{X}}(v) \sim \mathcal{N}(\mu_{\tilde{x}}, \sigma_{\tilde{x}}^2)$, where $\mu_{\tilde{x}}= f_X(v)$, $\sigma_{\tilde{x}}^2=\frac{q(1-q)}{n(p-q)^2} + \frac{ f_X(v) (1-p-q)}{n(p-q)}$, and $f_X(v)$ is the true frequency of $v$ in genuine data $X$.
\end{lemma}
\begin{proof}
	Please refer to our technical report \cite{sunxy}.
\end{proof}

\vspace{0.3em} \noindent \textbf{Model the Distribution of $\tilde{f}_{Z}(v)$.} 
Note that Lemmas~\ref{lmm_y_normal} and~\ref{lmm_x_normal} state that the aggregated frequency of each item in the data, sent to the server by either malicious or genuine users, approximates a specific normal distribution. Thus, the distribution of $\tilde{f}_{Z}(v)$ can be expressed jointly by the distributions of $\tilde{f}_{\tilde{X}}(v)$ and $\tilde{f}_{Y}(v)$. The following theorem establishes the asymptotic distribution of $\tilde{f}_Z(v)$.

\begin{theorem}\label{thm_z_normal_1}
	The asymptotic distribution of $\tilde{f}_Z(v)$ is $\mathcal{N}(\mu_{z}, \sigma_{z}^2)$, i.e.,  $\lim\limits_{n, m \to \infty} \tilde{f}_{Z}(v) \sim \mathcal{N}(\mu_{z}, \sigma_{z}^2)$, where $\mu_z = \frac{1}{1+\eta}\mu_{\tilde{x}}+\frac{\eta}{1+\eta}\mu_{y}$,  $\sigma_z^2 =\frac{1}{(1+\eta)^2} \sigma_{\tilde{x}}^2 +\frac{\eta^2}{(1+\eta)^2}\sigma_{y}^2$, and $\eta$ denotes the ratio of the number of malicious users to the number of genuine users, i.e., $\eta=\frac{m}{n}$.
\end{theorem}
\begin{proof}
	Please refer to our technical report \cite{sunxy}.
\end{proof}

\subsubsection{Estimator Construction}
Our analytic framework reveals the relationship of the distributions of $\tilde{f}_{\tilde{X}}(v)$, $\tilde{f}_{Y}(v)$, and $\tilde{f}_Z(v)$. On this basis, we propose the genuine frequency estimator to recover $\tilde{f}_{\tilde{X}}(v)$ from $\tilde{f}_Z(v)$ as follows.
\begin{equation}\label{equ_estimator}
	\tilde{f}_{\tilde{X}}(v) = (1+\eta)\tilde{f}_Z(v) -\eta \tilde{f}_{Y}(v).
\end{equation}
This estimator shows that the server with $\tilde{f}_Z(v)$ can recover $\tilde{f}_{\tilde{X}}(v)$ by removing $\tilde{f}_{Y}(v)$. Theorems \ref{thm_fx_unbias_1} and \ref{thm_fx_var_1} analyze the expectation and variance of our estimator in Equation (\ref{equ_estimator}). As we will show in experiments, setting a larger $\eta$ is sufficient to ensure good performance of the recovered aggregated frequencies, even though the server does not know the true value of $\eta$, 
\begin{theorem}\label{thm_fx_unbias_1}
	The estimator in Equation (\ref{equ_estimator}) is approximately unbiased, i.e., $\lim\limits_{n, m \to \infty} \mathbf{E}[	\tilde{f}_{\tilde{X}}(v)] = f_X(v)$, where $f_X(v)$ is the true frequency of $v$ in the genuine data $X$.
\end{theorem}
\begin{proof}
	Please refer to our technical report \cite{sunxy}.
\end{proof}

\begin{theorem}\label{thm_fx_var_1}
	The approximate variance of the estimator (Equation (\ref{equ_estimator})) is $\sigma_{\tilde{x}}^2$, where $\sigma_{\tilde{x}}$ can be obtained from Lemma \ref{lmm_x_normal}.
\end{theorem}
\begin{proof}
	Please refer to our technical report \cite{sunxy}.
\end{proof}

\subsection{Malicious Frequency Learning}\label{sec_malicious_frequecny_learning}
This subsection corresponds to Step 2 in LDPRecover. 

\vspace{0.3em} \noindent \textbf{Adaptive Attack.} Here we introduce an adaptive poisoning attack that unifies existing poisoning attacks~\cite{chen2021private,cao2021data} on the LDP protocols. We observe that existing poisoning attacks can be characterized as the sampling of malicious data for each malicious user from an attacker-designed distribution. Different attacks utilize different attacker-designed distributions, leading to various attackers' objectives. Accordingly, the adaptive attack is designed as follows: the attacker initially establishes an attacker-designed (adaptive) distribution $P$ over the encoded domain $\tilde{D}$, and subsequently draws samples $v \in \tilde{D}$ from $P$ with probability $P(v)$. These samples are then employed as the crafted data for malicious users. For example, the malicious data in MGA~\cite{cao2021data} are essentially sampled from the attacker-designed distribution in which the sampling probabilities of all untarget items are $0$. 

\vspace{0.3em} \noindent \textbf{Approximate Summation of Malicious Frequencies.} Next, we estimate the statistics of malicious frequencies during an adaptive attack as an approximate substitute for \(\tilde{f}_{Y}(v)\), since the server is unaware of \(\tilde{f}_{Y}(v)\) in the genuine frequency estimator. Specifically, during the adaptive poisoning attack, we can derive the expected summation of malicious frequencies for all items, denoted by \(\mathbf{E}[\sum_{v \in D} \tilde{f}_{Y}(v)]\). 
\begin{equation} \label{equ_expect_summation_P} 
	\mathbf{E}[\sum\nolimits_{v \in D} \tilde{f}_{Y}(v)] =  \frac{\sum_{v \in D} f_{Y}(v)- qd}{p-q}   =  \frac{1 - qd}{p-q}
\end{equation}
where $f_{Y}(v)$ is the frequency of item $v$ in $Y$, $d$ is the size of $D$, and $p, q$ are perturbation probabilities of the LDP protocol. Note that the value of $\mathbf{E}[\sum_{v \in D} \tilde{f}_{Y}(v)] $ changes with varying parameters $p$ and $q$ in different LDP protocols. This is due to the fact that the data reported by malicious users bypasses the perturbation algorithm of the LDP protocol, but is subjected to the aggregation algorithm of the LDP protocol. Additionally, regardless of the attacker-designed distribution $P$, we always have $\sum_{v \in D} f_{Y}(v) = 1$. Consequently, we can use the expected summation to approximate the actual summation of malicious frequencies for all items, i.e., 
\begin{equation}\label{equ_estimate_sum_fy}
	\sum\nolimits_{v \in D} \tilde{f}_{Y}(v) \triangleq \mathbf{E}[\sum\nolimits_{v \in D} \tilde{f}_{Y}(v)] = \frac{1 - qd}{p-q}.
\end{equation}
These approximate summations provide the necessary information for recovering the genuine frequencies from the poisoned ones in the subsequent section.

\subsection{Genuine Frequency Recovery}\label{sec_genuine_frequency_recovery}
This subsection corresponds to Step 3 in LDPRecover. Intuitively, we note that the process of recovering genuine frequency is essentially solving a Constraint Inference problem. A natural solution is to recover genuine frequencies from the poisoned ones using Constrained Least Squares (CLS)~\cite{hay2009boosting}, where the constraints are the genuine frequency estimator in Step 1 and the estimated summations of malicious frequencies in Step 2. We also impose the publicly prior knowledge that all individual frequencies should be non-negative, and they sum up to one as the constraints to improve the accuracy \cite{wang2019consistent}.

Formally, given the poisoned frequency vector $\mathbf{\tilde{f}}_Z$, LDPRecover outputs the recovered frequency vector $\mathbf{f}'_{\tilde{X}}$ by solving the following problem:
\begin{align}
	\text{minimize:} \ \ &||\mathbf{f}'_{\tilde{X}} - \tilde{\mathbf{f}}_{\tilde{X}} ||_2 \nonumber \\
	\text{subject to:}\ \  &\forall_v f'_{\tilde{X}}(v) \geq 0  \label{st_leq_0} \\
	& \sum\nolimits_{v \in D} f'_{\tilde{X}}(v) = 1 \label{st_sum_to_1}\\
	&\tilde{f}_{\tilde{X}}(v) = (1+\eta)\tilde{f}_Z(v) - \eta \tilde{f}_Y(v) \label{st_genuine_frequency}  \\
	&\sum\nolimits_{v\in D} \tilde{f}_Y(v)  =	\frac{1- qd}{p-q} \label{st_sum_fy_D}
\end{align}
where the first two conditions come from the publicly prior knowledge, and the last two derive from Steps 1 \& 2 of LDPRecover. To solve this problem, we first use conditions (\ref{st_genuine_frequency}) and (\ref{st_sum_fy_D}) to estimate the genuine frequencies $\tilde{\mathbf{f}}_{\tilde{X}}$, and then refine $\mathbf{f}'_{\tilde{X}}$ from $\tilde{\mathbf{f}}_{\tilde{X}}$ under conditions (\ref{st_leq_0}) and (\ref{st_sum_to_1}).

\vspace{0.3em} \noindent \textbf{Estimating Genuine Frequencies.} Specifically, we consider two scenarios: non-knowledge scenario and partial-knowledge scenario. The former assumes that the server is unaware of the attack details, while the latter assumes that the server knows attacker-selected items in the targeted attack.

\vspace{0.3em} \noindent \textit{Non-Knowledge Recovery.} In the non-knowledge scenario, the server has no details about the attack. We divide the domain $D$ into two sub-domain $D_0 \subseteq D$ and $D_1 = D \setminus D_0$ such that $D_0 = \{v | \tilde{f}_Z(v) \leq 0\}$ and $D_1 = D \setminus D_0$. Intuitively, the poisoned frequencies of potential items subjected to poisoning attacks should be higher. Thus, the items in $D_1$ are considered to be potential items subject to poisoning attacks, and their malicious frequencies are assumed as uniform. That is, for any $ v \in D_0$, we assign $f'_{Y} (v) = 0$, while for any $v \in D_1$, we assign 
\begin{equation}\label{equ_estimate_fy_D}
	\tilde{f}'_Y(v)  = \frac{1}{|D_1|} \sum\nolimits_{v\in D} \tilde{f}_Y(v).
\end{equation}
where $\sum_{v\in D} \tilde{f}_Y(v)$ can be obtained from Equation (\ref{st_sum_fy_D}). By replacing $\tilde{f}_Y(v)$ with $\tilde{f}'_Y(v)$ in Equation (\ref{st_genuine_frequency}), we obtain the estimated genuine frequency for each $v \in D$ as follow.
\begin{equation}\label{est_non_genuine_frequency}  
	\tilde{f}_{\tilde{X}}(v) = (1+\eta)\tilde{f}_Z(v) - \eta \tilde{f}'_Y(v)
\end{equation}

\vspace{0.3em} \noindent \textit{Partial-Knowledge Recovery.} We observe that targeted attacks tend to significantly increase the frequencies of attacker-selected items, rendering these items as statistical outliers. There are many outlier detection techniques~\cite{kieu2018outlier,zhou2018non,su2019robust} available for inferring the target items by analyzing statistical anomalies in the frequency of each item in historical data. For example, these works can encode the historical frequencies of each item as a time series, fit a prediction model to the time series and predict current frequencies using past data. Aggregated frequencies of items that are very different from their predicted frequencies are identified as outliers (i.e., target items). Motivated by this, in the partial-knowledge scenario, we assume that the server has awareness of the items selected by the attacker. In what follows, we illustrate how to integrate the available attack details into LDPRecover to estimate more accurate genuine frequencies.

Specifically, let $\mathcal{T}$ denote the set of attacker-selected items. With $\mathcal{T}$, LDPRecover can update $\tilde{f}'_{Y}(v)$ in Equation (\ref{equ_estimate_fy_D}) as its more accurate version, denoted by $\tilde{f}^*_Y(v)$. Note that $\mathcal{T}$ affects the estimated result of $\mathbf{f}_{Y}$. Therefore, we first use $\mathcal{T}$ to obtain $\tilde{f}^*_{Y}(v)$ for all $v \in D$. In this step, we divide $D$ into two cases: $D' =  D \setminus \mathcal{T}, D''=D \cap \mathcal{T}$. For both cases, we have the following estimated results:

\begin{itemize}[leftmargin=*]
	\item $v \in D'$: As $D'$ does not include any target item in $\mathcal{T}$, the frequencies of $v$ in $D'$ in malicious data $Y$ are $0$, i.e., $P(v) = 0$ for all $v \in D'$. Following Equation (\ref{equ_expect_summation_P}), the approximate summation of malicious frequencies for all items in $D'$ can be represented as follows.
	\begin{equation}
		\sum\nolimits_{v \in D'} \tilde{f}_Y(v) \triangleq \frac{\sum_{v \in D'} P(v) - qd}{p-q} = -\frac{qd}{p-q}
	\end{equation}
	
	\item $v \in D''$: In this case, $v$ is the attacker-selected item in $\mathcal{T}$. With $\sum_{v \in D'} \tilde{f}_Y(v)$, the approximate summation of malicious frequencies for all items in $D''$ can be computed as follows.
	\begin{align}
		\sum\nolimits_{v \in D''} \tilde{f}_Y(v) &= \sum\nolimits_{v \in D} \tilde{f}_Y(v) - \sum\nolimits_{v \in D'} \tilde{f}_Y(v)  
	\end{align}
	where $\sum_{v \in D} \tilde{f}_Y(v)$ is obtained from Equation (\ref{st_sum_fy_D}). 
\end{itemize}

As we do not know the weights among the attacker-selected items, we again assume that the frequencies of such items are uniformly distributed. Putting things together, we have the estimate the new malicious frequencies $\tilde{f}^*_Y(v)$ as follows.
\begin{equation}\label{equ_estimate_fy*_D}
	\tilde{f}^*_Y(v) = \begin{cases}
		-\frac{qd}{|D'|(p-q)}, & v \in D' \\
		\frac{1}{|D''|}(\sum\limits_{v \in D} \tilde{f}_Y(v) - \sum\limits_{v \in D'} \tilde{f}_Y(v) ), & v \in D''
	\end{cases}
\end{equation}

Next, we can update Equation (\ref{est_non_genuine_frequency}) with $\tilde{f}^*_Y(v)$ to obtain the recovered frequency as follows.
\begin{equation}\label{est_partial_genuine_frequency}  
	\tilde{f}_{\tilde{X}}(v) = (1+\eta)\tilde{f}_Z(v) - \eta \tilde{f}^*_{\tilde{X}}(v)
\end{equation}

Intuitively, such partial knowledge can improve the accuracy of genuine frequencies, thereby enabling LDPRecover to recover a more accurate aggregated frequency. Our experiments in Section \ref{sec:experimental_results} confirm this intuition.

\vspace{0.3em} \noindent \textbf{Refining Recovered Frequencies.} With $\tilde{\mathbf{f}}_{\tilde{X}}$, we use the KKT condition~\cite{wang2019consistent,karush1939minima, kuhn2014nonlinear} to solve this problem under conditions (\ref{st_leq_0}) and (\ref{st_sum_to_1}). Specifically, the optimization target can be augmented by the following equations:
\begin{align}
	\text{minimize:} \ \ & \sum\nolimits_{v \in D}(f'_{\tilde{X}}(v) - \tilde{f}_{\tilde{X}}(v) )^2  + a + b\nonumber \\
	\text{where:}\ \  &\forall_v f'_{\tilde{X}}(v) \geq 0, \sum\nolimits_{v} f'_{\tilde{X}}(v) = 1 \nonumber \\ 
	& a = \mu \sum_{v}f'_{\tilde{X}}(v)  \nonumber \\
	& b = \sum\nolimits_{v}\lambda_v f'_{\tilde{X}}(v), \forall_v: \lambda_v \times f'_{\tilde{X}}(v) = 0 \nonumber
\end{align}
Note that $a = \mu$ is a constant, and $b=0$. Therefore, the conditions of the minimization objective are unchanged. As the target is convex, we can find the minimum by deriving the derivative for each variable $f'_{\tilde{X}}(v)$:
\begin{equation}
	\begin{split}
		&\frac{\partial \sum_{v}(f'_{\tilde{X}}(v) - \tilde{f}_{\tilde{X}}(v) )^2 + a +b}{\partial f'_{\tilde{X}}(v)} = 0 \\
		\longrightarrow & f'_{\tilde{X}}(v) =  \tilde{f}_{\tilde{X}}(v) - \frac{1}{2}(\mu + \lambda_v) \label{equ_ff_x}
	\end{split}
\end{equation}
Then, we re-divide the domain $D$ into two sub-domain $D^* \subseteq D$ and $D_* = D \setminus D^{*}$ such that $\forall v \in D^*, f'_{\tilde{X}} (v) = 0$ and $\forall v \in D_*, f'_{\tilde{X}}(v) > 0 \wedge \lambda_v = 0$.  For all $v \in D$, we have 
\begin{equation}
	1 = \sum_{v \in D^*} f'_{\tilde{X}}(v) + \sum_{v \in D_*} f'_{\tilde{X}}(v) =  \sum_{v \in D_*} \tilde{f}_{\tilde{X}}(v)- \frac{|D_*|\mu}{2} \label{equ_sum_up_1}
\end{equation}
From this, we can derive $\mu$ as follows.
\begin{equation}
	\mu = \frac{2}{|D_*|}( \sum\nolimits_{v \in D_*}\tilde{f}_{\tilde{X}}(v)- 1)
\end{equation}
By plugging $\mu$ into Equation (\ref{equ_ff_x}), for all $v \in D_*$ we have $f'_{\tilde{X}}(v)$ as follows:
\begin{equation}\label{equ_estimate_fx_D1__}
	f'_{\tilde{X}}(v) =  \tilde{f}_{\tilde{X}}(v) - \frac{1}{|D_*|}( \sum\nolimits_{v \in D_*} \tilde{f}_{\tilde{X}}(v)- 1).
\end{equation}

\vspace{0.3em} \noindent \textbf{Implementation of LDPRecover.} We provide the pseudo-code for LDPRecover without prior knowledge in Algorithm \ref{Alg_LDPRecover}. Specifically, we first estimate the genuine frequencies (lines 1-4): we initialize $D_0, D_1$, compute the malicious frequencies $\tilde{f}'_Y(v)$ for each $v \in D$ and estimate the genuine frequencies $f'_{\tilde{X}}(v)$ for each $v \in D$. Then, we through an iterative process of finding $D^*$ to refine the recovered frequencies based on the genuine frequencies (lines 5-11): we initiate $D^*=\emptyset$ and $D_*=D$, and then iteratively test whether $f'_{\tilde{X}}(v)$ for all $v \in D_*$ are positive. In each iteration, we move $v$ from $D_*$ to $D^*$ if $f'_{\tilde{X}}(v)$ is negative. The iterative procedure is repeated until $f'_{\tilde{X}}(v)$ for all $v \in D_*$ are non-negative. In particular, Algorithm \ref{Alg_LDPRecover} becomes the pseudo-code for LDPRecover with partial knowledge by replacing $\tilde{f}'_Y(v)$ and Equations (\ref{equ_estimate_fy_D}) and (\ref{est_non_genuine_frequency}) with $\tilde{f}^*_Y(v)$ and Equations (\ref{equ_estimate_fy*_D}) and (\ref{est_partial_genuine_frequency}).

\begin{algorithm}[t]
	\caption{LDPRecover}
	\label{Alg_LDPRecover}
	\hspace*{0.02in} \raggedright  {\bf Input:} {Poisoned frequencies $\mathbf{\tilde{f}}_Z$, estimated sum of malicious frequencies $\sum_{v \in D} \tilde{f}_Y(v)$, $\eta$} \\
	\hspace*{0.02in} \raggedright {\bf Output:} Recovered frequencies $\mathbf{f}'_{\tilde{X}}$ 
	\begin{algorithmic}[1]
		\State {\small // Estimate genuine frequencies}
		\State {\small Initialize $D_0 = \{v | \tilde{f}_Z(v) \leq 0\}$ and $D_1 = D \setminus D_0$}
		\State  {\small Compute $\tilde{f}'_Y(v)$ for all $v \in D$ according to Equation (\ref{equ_estimate_fy_D})}
		\State  {\small Estimate $\tilde{f}_{\tilde{X}}(v)$ for all $v \in D$ according to Equation (\ref{est_non_genuine_frequency})}
		\State {\small // Refine recovered frequencies}
		\State {\small Initialize $D^*=\emptyset$ and $D_*=D$}
		\State  {\small Initialize $f'_{\tilde{X}}(v)$ for all $v \in  D_*$ according to Equation (\ref{equ_estimate_fx_D1__})}
		\While{\small for all $v \in D_*, \min\{f'_{\tilde{X}}(v)\} < 0$}
		\State  {\small Move $v \in D_*$ from $D_*$ to $D^*$ if $f'_{\tilde{X}}(v)< 0$ }
		\State  {\small Update $f'_{\tilde{X}}(v)$ for all $v \in  D_*$ according to Equation (\ref{equ_estimate_fx_D1__})}
		\EndWhile
		\State \Return $\mathbf{f}'_{\tilde{X}}$ 
	\end{algorithmic}
\end{algorithm}

\subsection{Approximation Error of LDPRecover}\label{sec_error_analysis}
LDPRecover is built on the genuine frequency estimator, which assumes the server receives sufficient reports from genuine and malicious users. When this assumption no longer holds, the \textit{Lindeberg–L\'evy} CLT theorem provides an asymptotic approximation of the genuine frequency estimator. To understand the gap between the asymptotic estimator and the true one, we analyze the approximation error of $\tilde{f}_{\tilde{X}}(v)$ and $\tilde{f}_{Y}(v)$ in our estimator in terms of the numbers of genuine and fake users, i.e., $n$ and $m$. 

Specifically, for $\tilde{f}_{Y}(v)$, suppose its true \textit{probability density function} (pdf) is $\tilde{\theta}_{Y,v}$, then its \textit{cumulative distribution function} (cdf) would be $\tilde{\Theta}_{Y,v} (w) = \int_{-\infty}^{w}\tilde{\theta}_{Y,v}(\tilde{f}_{Y}(v))d(\tilde{f}_{Y}(v))$. According to Lemma \ref{lmm_y_normal}, the approximated \textit{pdf} of $\tilde{f}_{Y}(v)$ is $\hat{\theta}_{Y,v} (\tilde{f}_{Y}(v))=\frac{1}{\sqrt{2\pi}\sigma_y}\exp(-\frac{(\tilde{f}_{Y}(v)-\mu_y)^2}{2 \sigma_y^2})$, and its \textit{cdf} is $\hat{\Theta}_{Y,v}(w)= \int_{-\infty}^{w}\hat{\theta}_{Y,v} (\tilde{f}_{Y}(v))d(\tilde{f}_{Y}(v))$. Then we have:
\begin{theorem}\label{thm_approximate_y_error}
	For $\tilde{f}_{Y}(v)$, its true \textit{cdf} $\tilde{\Theta}_{Y,v} (w)$ and approximated \textit{cdf} $\hat{\Theta}_{Y,v}(w)$ differ by no more than $\frac{0.33554(g_y + 0.415\sigma_y^3)}{\sigma_y^3\sqrt{m}}$, where $g_y = \mathbf{E}[|\tilde{f}_{Y}(v) - \mu_y|^3]$, and $\mu_y, \sigma_y$ can be obtained from Lemma \ref{lmm_y_normal}.
\end{theorem}
\begin{proof}
	Please refer to our technical report \cite{sunxy}.
\end{proof}

Similar to $\tilde{f}_{Y}(v)$, we have the error bound of $\tilde{f}_{\tilde{X}}(v)$ as follows.
\begin{theorem}\label{thm_approximate_x_error}
	For $\tilde{f}_{\tilde{X}}(v)$, its true \textit{cdf} $\tilde{\Theta}_{\tilde{X},v} (w)$ and approximated \textit{cdf} $\hat{\Theta}_{\tilde{X},v}(w)$ differ by no more than $\frac{0.33554(g_x + 0.415\sigma_x^3)}{\sigma_x^3\sqrt{n}}$, where $g_x = \mathbf{E}[|\tilde{f}_{\tilde{X}}(v) - \mu_x|^3]$, and $\mu_x, \sigma_x$ can be obtained from Lemma \ref{lmm_x_normal}.
\end{theorem}

In Theorems \ref{thm_approximate_y_error} and \ref{thm_approximate_x_error}, the gap between asymptotic gain and true one can be defined as functions of $n$ and $m$. Therefore, the speed of convergence rate in the asymptotic distributions of $\tilde{f}_{Y}(v)$ and $\tilde{f}_{\tilde{X}}(v)$ are at least on the order of $\frac{1}{\sqrt{n}}$ and $\frac{1}{\sqrt{m}}$, respectively. That is, the approximation error is still tolerable even if the number of reports is insufficient. 

\section{Evaluation}\label{Sec_evaluate}

\subsection{Experimental Setup}\label{sec:experimental_setup}
\subsubsection{Datasets}
We evaluate our proposed LDPRecover on two real-world datasets, i.e., IPUMS~\cite{ipums} and Fire~\cite{fire}. 

\vspace{0.3em} \noindent \textbf{IPUMS~\cite{ipums}:} The IPUMS dataset encompasses U.S. census data collected over several years. For this study, we have chosen to focus on the most recent data from 2017 and have identified the attribute of ``city'' as the item held by each user. This yields a total of 102 items across 389,894 users.

\vspace{0.3em} \noindent \textbf{Fire~\cite{fire}:} The Fire dataset, collated by the San Francisco Fire Department on January 16, 2023, documents details pertaining to service calls. We refine these records based on call type and utilize data corresponding to the ``Alarms" category. The ``unit ID'' is considered as the item held by each user, culminating in a total of 490 items and 667,574 users.

\subsubsection{LDP Settings}
We consider three popular LDP protocols for frequency estimation, namely GRR, OUE, and OLH, the details of which are given in Section \ref{sec_LDP_protocols}. We set the default parameters of the LDP protocols as $\epsilon=0.5$ and $g = \lceil e^{\epsilon}+1 \rceil$.

\subsubsection{Attack Settings}
We consider two state-of-the-art poisoning attacks to LDP, namely \textbf{Manip}~\cite{cheu2021manipulation} and \textbf{MGA}~\cite{cao2021data}, as well as our proposed \textbf{adaptive attack (AA)} already described in Section~\ref{sec_malicious_frequecny_learning}. For Manip, we first sample a malicious data domain $H$ from the data domain $D$, and then draw uniform samples (malicious data) from $H$. For MGA, we randomly select target items and draw samples from the attacker-designed distribution, where the sampling probabilities of all untarget items are $0$ and that of the target items are $1/r$ ($r$ is the number of target items). For AA, we randomly generate the attacker-designed distribution, then draw samples from this distribution and use the samples as the malicious users' data.

The parameters involved in the LDP protocols and attacks are $\beta = \frac{m}{n+m}$ (the fraction of malicious users) and $r$ (the number of target items in MGA). We set the default values of these parameters as $\beta=0.05$ and $r=10$. 

\subsubsection{Recovery Settings}
For ease of presentation, we use LDPRecover and LDPRecover$^*$ to denote the versions of LDPRecover operating without and with partial knowledge of the items selected by the attacker, respectively. In LDPRecover, the server aggregates item frequencies without any detail of the attacks. Conversely, LDPRecover$^*$ operates under the assumption that the server is aware of the attacker-selected items. In the context of MGA, these items are explicitly identified as target items, while in AA, they are the items that exhibit the top-$r/2$ frequency increase following the attack.

The parameter involved in LDPRecover and LDPRecover$^*$ is $\eta=\frac{m}{n}$ (the ratio of the number of malicious users to the number of genuine users). Since, in practice, the server does not know what the real value of $\eta$ is, we set a large $\eta=0.2$ by default in the experiments, which is a value larger than the real value (i.e., $\eta \gg \frac{\beta}{1-\beta}$). We also explore the impact of $\eta$ on recovered results from the poisoning attacks in Section \ref{sec_experiment_various_parameters}, where the range of $\eta$ is $\eta \in [0.01, 0.4]$.

\subsubsection{Comparison Methods}
To the best of our knowledge, LDPRecover is the only method that recovers the accurate aggregated frequencies from the poisoned ones, whether or not the details of the attacks are known.
To achieve a fair comparison, we incorporate partial knowledge (in Section~\ref{sec_genuine_frequency_recovery}) to adapt the detection method. Specifically, Detection identifies users as malicious if their reported data matches the target items.

\begin{figure}[htbp]
	\centerline{\includegraphics[width=0.9\linewidth]{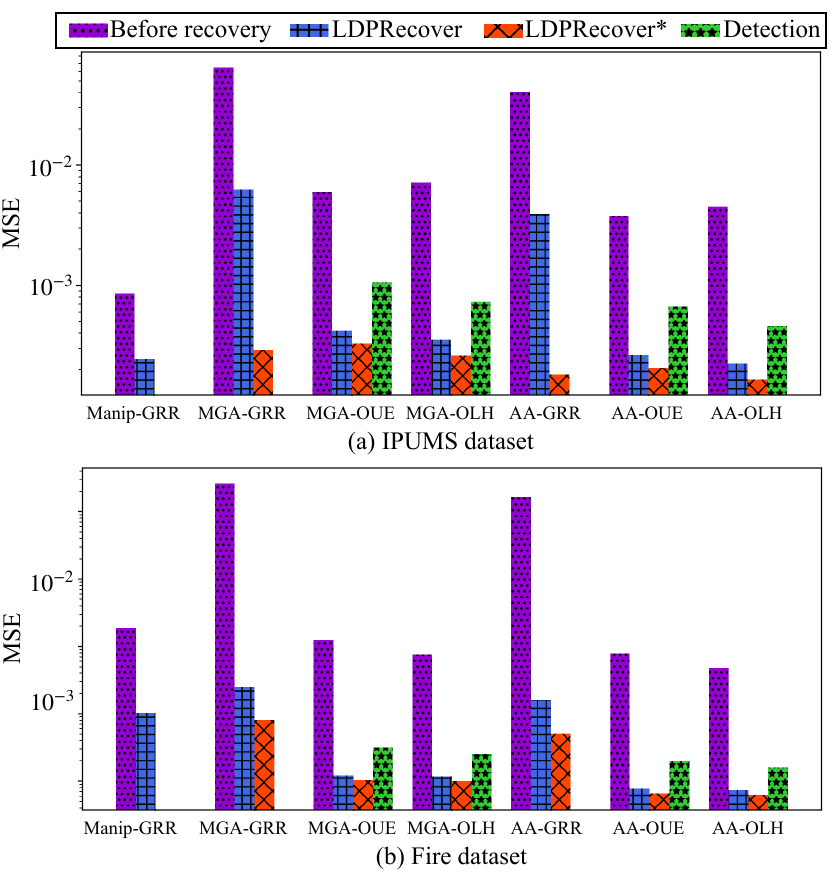}}
	\caption{The mean squared error (MSE) of LDPRecover and LDPRecover$^*$ for two datasets, three LDP protocols, and three attacks. ``-Manip'', ``-MGA'', and ``-AA'' represents the results for recovery from Manip, MGA, and AA, respectively.}
	\label{fig_hist_for_overall_mse}
\end{figure} 

\begin{figure}[htbp]
	\centerline{\includegraphics[width=0.9\linewidth]{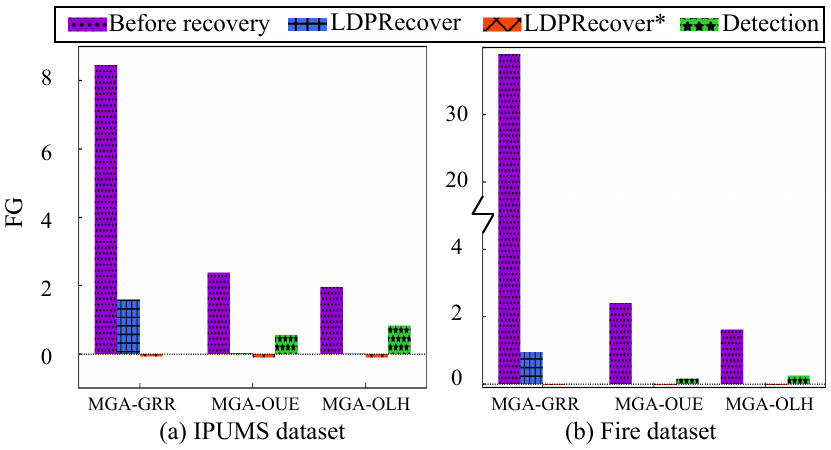}}
	\caption{The frequency gain (FG) of LDPRecover and LDPRecover$^*$ for two datasets, three LDP protocols, and three attacks. ``-Manip'', ``-MGA'', and ``-AA'' represents the results for recovery from Manip, MGA, and AA, respectively.}
	\label{fig_hist_for_overall_fg}
\end{figure} 

\begin{figure}[htbp]
	\centerline{\includegraphics[width=\linewidth]{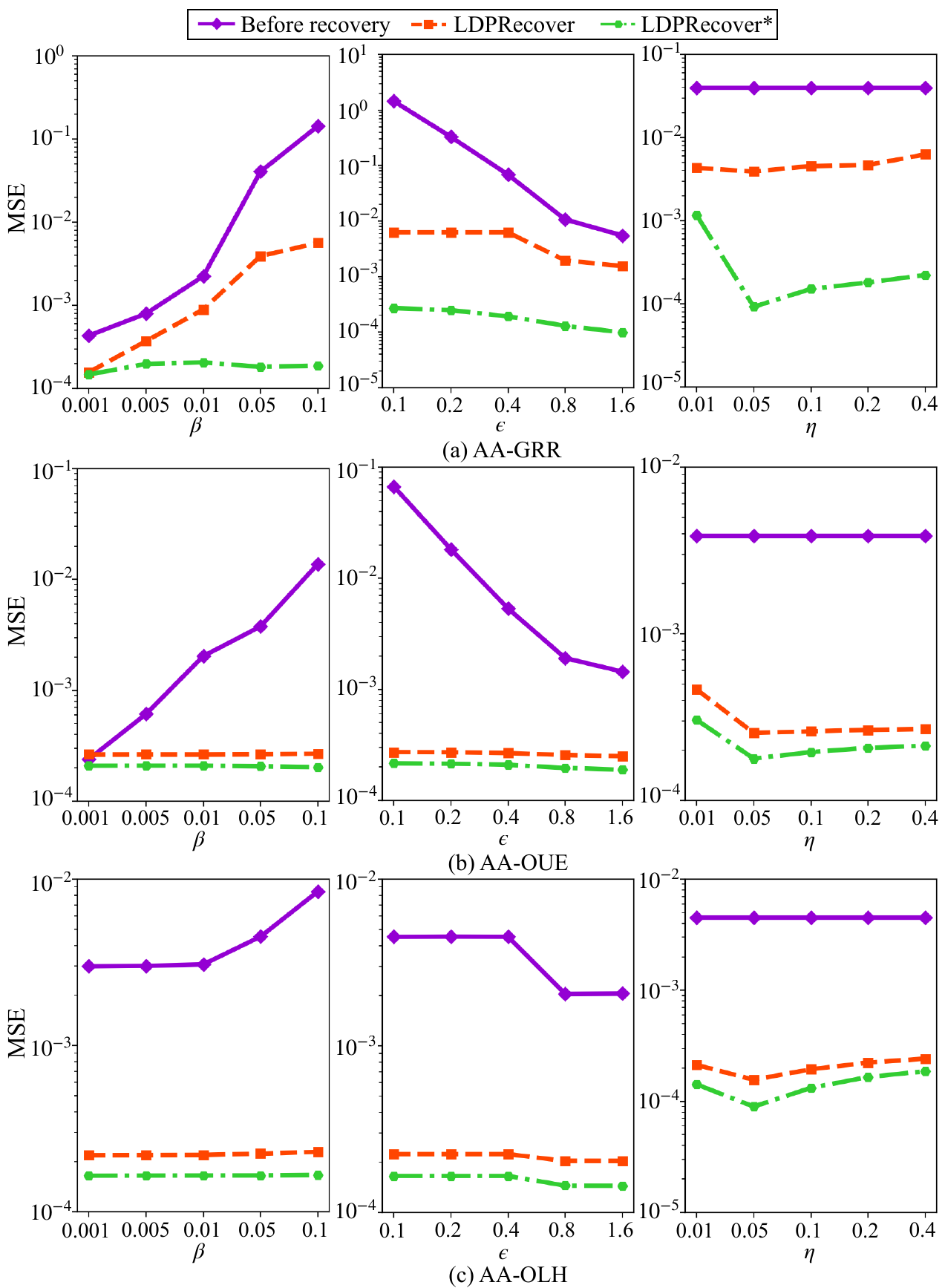}}
	\caption{Impact of differential parameters ($\beta, \epsilon, \eta$) on recovery from AA on IPUMS dataset in terms of MSE.}
	\label{fig_mse_for_ipums_aa}
\end{figure} 

\begin{figure}[htbp]
	\centerline{\includegraphics[width=0.9\linewidth]{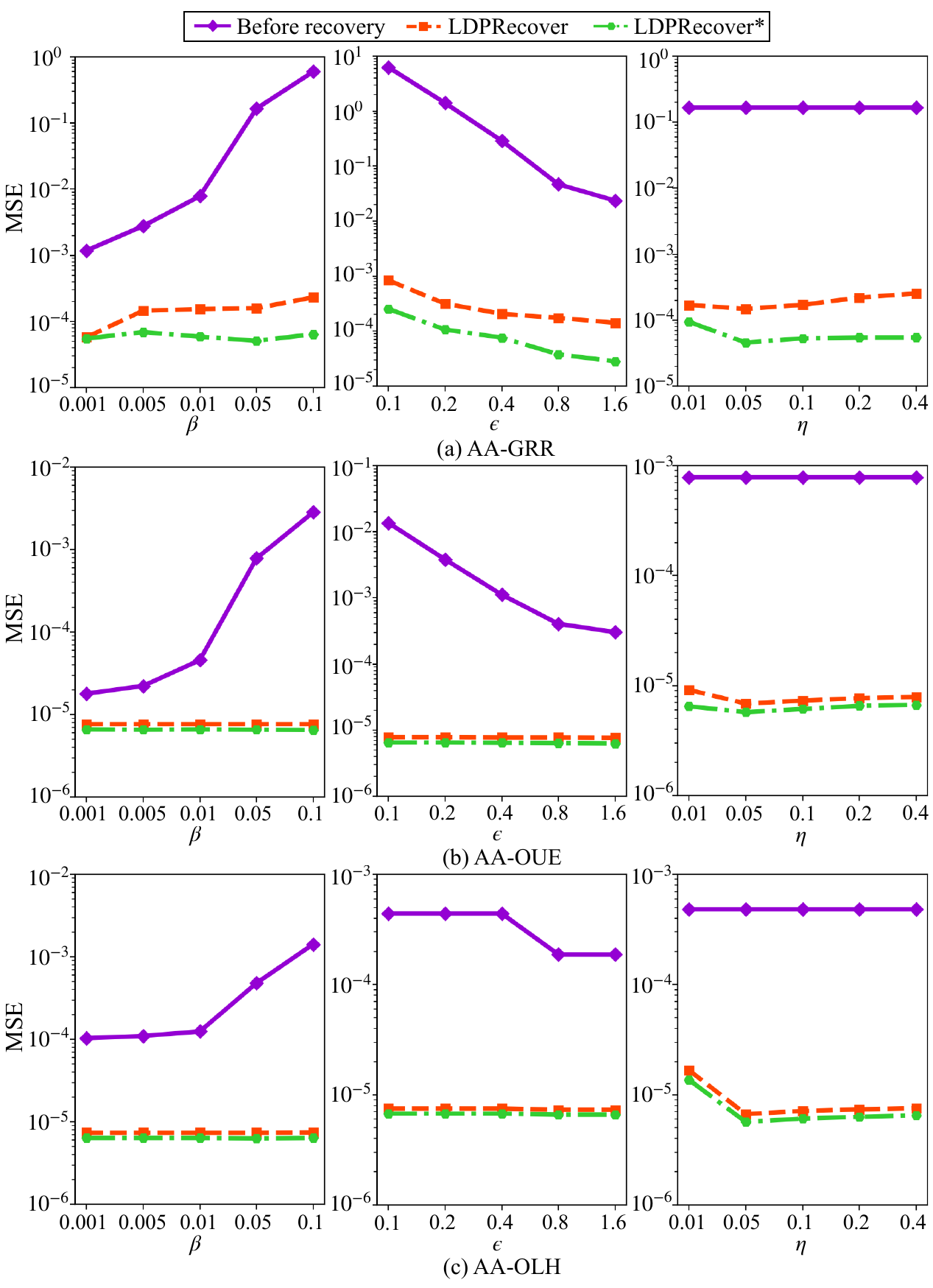}}
	\caption{Impact of differential parameters ($\beta, \epsilon, \eta$) on recovery from AA on fire dataset in terms of MSE.}
	\label{fig_mse_for_fire_aa}
\end{figure} 

\subsection{Evaluation Metrics}\label{sec:evaluation_metrics}
We adopt mean squared error (MSE) and frequency gain (FG) as evaluation metrics. We define them as follows.

\vspace{0.3em} \noindent \textbf{Mean Squared Error (MSE).} Given original frequencies and (recovered or poisoned) aggregated frequencies, we use the MSE to evaluate the average error of frequencies for all items. Specifically, 
\begin{align}
	\text{MSE} = \frac{1}{d} \sum\nolimits_{v \in D} ( \tilde{f}_{X}(v) - \tilde{f}^*_Z(v))^2,
\end{align}
where $d$ is the size of the domain $D$, $ \tilde{f}_{X}(v)$ is the original frequency of $v$ in genuine data $X$, and $\tilde{f}^*_Z(v)$ could be the recovered or poisoned frequency of $v$. 

\vspace{0.3em} \noindent \textbf{Frequency Gain (FG).} For targeted poisoning attacks, we follow \cite{cao2021data} to evaluate the recovery methods via FG. Specifically, given a set of target items $\mathcal{T}$, FG is defined as the sum of frequency gain of each target item $t \in \mathcal{T}$, i.e., 
\begin{align}
	\text{FG} = \sum\nolimits_{t \in \mathcal{T}}(\tilde{f}_{\tilde{X}}(t) - \tilde{f}^*_Z(t)),
\end{align}
where $\tilde{f}_{\tilde{X}}(v)$ is the genuine frequency of $v$ aggregated from genuine data $X$ using LDP protocols. 

We say a recovery method is more accurate and effective if the recovered frequencies have a smaller MSE and FG: smaller MSE means better accuracy, and smaller FG implies better counter targeted attacks. Since the frequency recovery process involves randomness, we average the results over $10$ trials to compute the MSE and FG in the experiments.

\subsection{Experimental Results}\label{sec:experimental_results}
\begin{figure}[tb]
	\centerline{\includegraphics[width=\linewidth]{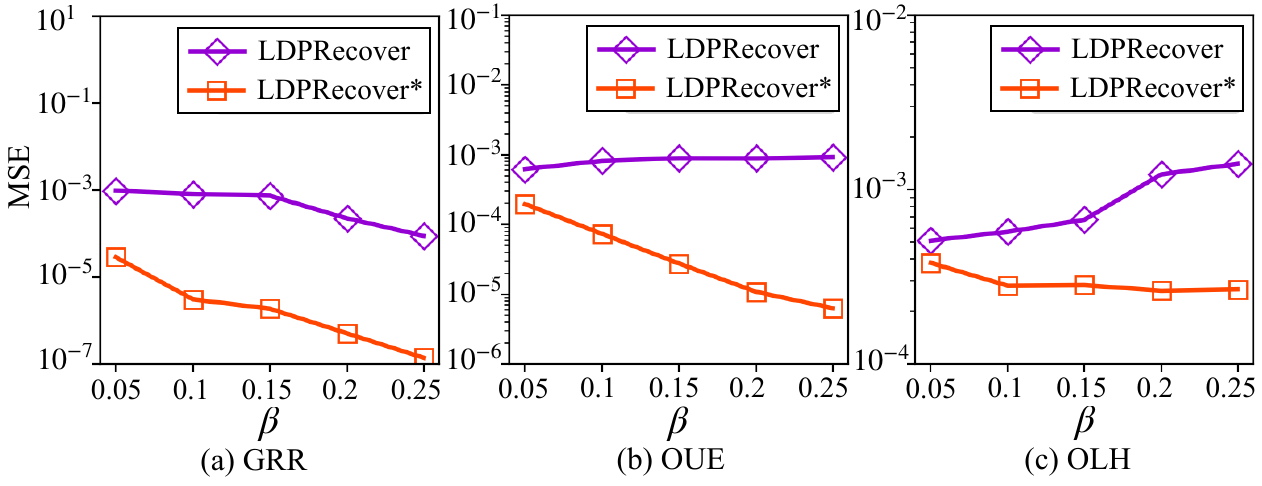}}
	\caption{MSE between LDPRecover and LDPRecover$^*$ estimated malicious frequencies and true malicious frequencies (IPUMS).}
	\label{fig_mal_fre-beta}
\end{figure}

Figures \ref{fig_hist_for_overall_mse} and \ref{fig_hist_for_overall_fg} show the MSE and FG of LDPRecover and LDPRecover$^*$ for the two datasets, three LDP protocols, and three attacks. From these figures, we have the following observations:

\vspace{0.3em} \noindent \textbf{Both LDPRecover and LDPRecover$^*$ are Accurate and Widely Applicable.} Both LDPRecover and LDPRecover$^*$ can recover the accurate aggregated frequencies from the poisoned ones caused by Manip, MGA, and AA, even if the default $\eta=0.2$ significantly exceeds the actual ratio $\frac{\beta}{1-\beta} = 0.052$. 
Notably, LDPRecover$^*$ consistently performs best when countering MGA, i.e., it achieves a lower MSE than LDPRecover. Specifically, both LDPRecover and LDPRecover$^*$ recover aggregated frequencies by subtracting the malicious frequencies from the poisoned frequencies, i.e., the more accurate the estimated malicious frequencies, the more accurate the recovered frequencies are. Based on this, while LDPRecover estimates these malicious frequencies based solely on the LDP protocol's properties, LDPRecover$^*$ refines these estimations by incorporating information about the attacker-selected items, thus enhancing accuracy. To confirm this, we evaluated the MSEs of the malicious frequencies estimated by LDPRecover and LDPRecover$^*$ versus the true malicious frequencies. The experimental results in Figure \ref{fig_mal_fre-beta} show that LDPRecover$^*$ estimates malicious frequencies more accurately than LDPRecover. Consequently, LDPRecover$^*$ surpasses LDPRecover in recovering the aggregated frequencies by estimating more accurate malicious frequencies. In other words, the prior knowledge of attacker-selected items introduced in LDPRecover$^*$ helps to improve the accuracy of the recovered frequency. In addition, we note that LDPRecover and LDPRecover$^*$ outperform Detection in all cases. This is because Detection removes all users with the target items indiscriminately, such that the genuine users with the target items are incorrectly removed.

\vspace{0.3em} \noindent \textbf{Both LDPRecover and LDPRecover$^*$ reduce FG of Targeted Attacks.} LDPRecover and LDPRecover$^*$ can effectively defend against targeted attacks. First, LDPRecover can significantly reduce the FG of targeted attacks, especially to  almost $0$ in most cases. Second, LDPRecover$^*$ even reduces the FG to a negative value (FG$<0$), which means that the recovered frequencies of the target items are even small than the ones before the attack. Third, Detection is much less effective in dealing with targeted attacks than LDPRecover, not to mention LDPRecover$^*$. This performance gap comes from the same reason: Detection brutally removes all users with target items. In contrast, LDPRecover$^*$, which has the same prior knowledge as Detection, can obtain more accurate malicious data, as shown in Figure \ref{fig_mal_fre-beta}. This allows LDPRecover$^*$ to finely deduct malicious data from poisoned data, gaining a lower FG.

\subsection{Effectiveness of Various Parameters}\label{sec_experiment_various_parameters}
In this subsection, we aim to study the impact of different parameter settings (i.e., $\beta, \epsilon$, and $\eta$) on the recovery results of our recovery methods from adaptive attacks. In particular, $\beta, \epsilon$ are the parameters of the poisoning attacks, while $\eta$ is the parameter of our recovery methods. Specifically, we vary one parameter while keeping the others ﬁxed to their default values to investigate its impact on our recovery methods, where the range of these parameters are $\beta = 0$ (i.e., LDPRecover executes on the unpoisoned data), $ \beta \in [0.001, 0.1]$, $\epsilon \in [0.1, 1.6]$, and $\eta \in [0.01, 0.4]$. 

\begin{table}[t]
	\centering
	\renewcommand\arraystretch{1.2}
	\caption{MSE of the LDPRecover execution on the unpoisoned frequencies}
	\scalebox{1.1}{\begin{tabular}{ccccc}
			\Xhline{1pt}
			\multirow{2}*{\centering LDP} &
			\multicolumn{2}{c}{IPUMS} &
			\multicolumn{2}{c}{Fire} \\
			\Xcline{2-5}{0.4pt}
			& {Before-Rec} & {After-Rec} & {Before-Rec} & {After-Rec} \\ 
			\Xhline{1pt}
			\multirow{1}*{\textit{GRR}}	&5.89E-4 &5.31E-4 &1.68E-3&3.62E-5\\
			\multirow{1}*{\textit{OUE}}	&3.81E-5 &5.33E-4 &2.93E-5&3.64E-5\\
			\multirow{1}*{\textit{OLH}}	&1.21E-6 &5.30E-4 &6.87E-7&3.63E-5\\
			\Xhline{1pt}						  		
	\end{tabular}}
	\label{tab_results_for_unpoisoned_frequency}
\end{table}

\vspace{0.3em} \noindent \textbf{Impact of the fraction of malicious users $\beta$.} The first column of Figures \ref{fig_mse_for_ipums_aa} and \ref{fig_mse_for_fire_aa} shows the impact of $\beta$ on recovering frequencies from AA over IPUMS and Fire datasets, respectively. We observe that when defensing against poisoning attacks, LDPRecover and LDPRecover$^*$ can recover the accurate aggregated frequencies under various $\beta$. Moreover, Table \ref{tab_results_for_unpoisoned_frequency} shows the MSE of the LDPRecover execution on the unpoisoned frequencies, where Before-Rec and After-Rec denote the MSE of unpoisoned frequencies and the MSE of the frequencies recovered by LDPRecover on the unpoisoned frequencies, respectively. From this table, we observe that when LDPRecover executes on the unpoisoned frequencies, it can improve the accuracy of the frequencies from GRR and reduce the accuracy of the frequencies from OLH and OLH. This is because when LDPRecover executes on the unpoisoned frequencies from OUE and OLH, it may remove frequencies that should not be removed, thus reducing accuracy. 


\vspace{0.3em} \noindent \textbf{Impact of the privacy budget of the LDP protocols $\epsilon$.} The second column of Figures \ref{fig_mse_for_ipums_aa} and \ref{fig_mse_for_fire_aa} shows the impact of the privacy budget on recovering frequencies from AA over IPUMS and Fire datasets, respectively. We observe that regardless of $\epsilon$, LDPRecover and LDPRecover$^*$ are both effective in recovering accurate aggregated frequencies. In particular, for LDPRecover$^*$, the MSE remains low and stable under in all cases; for LDPRecover, the MSE may decreases or remains stable as $\epsilon$ grows. The discrepancy arises because LDPRecover$^*$ exploits known details of the poisoning attack, resulting in a stable and accurate estimation. In contrast, LDPRecover utilizes the LDP protocol's properties for estimation, rendering it susceptible to fluctuations caused by $\epsilon$.

\vspace{0.3em} \noindent \textbf{Impact of the ratio of the number of malicious users to the number of genuine users $\eta$.} The third column of Figures \ref{fig_mse_for_ipums_aa} and \ref{fig_mse_for_fire_aa} shows the effect of $\eta$ on LDPRecover and LDPRecover$^*$. We observe that LDPRecover and LDPRecover$^*$ perform optimally when $\eta$ closely matches $\beta$. This is because a more accurate $\eta$ will enable our methods to estimate genuine frequencies more accurately. Furthermore, they still maintain effectiveness even there is a deviation between $\eta$ and $\beta$. This is because, even if the estimated genuine frequencies are moderately accurate, our method can still refine accurate recovered frequencies from the estimated genuine frequencies using public available knowledge (i.e., Equations (\ref{st_leq_0}) and (\ref{st_sum_to_1})).	For example, Figure \ref{fig_mse_for_ipums_aa} (a) illustrates that with $\beta=0.05$ and $\eta=0.4$, LDPRecover significantly outperforms the poisoned data, with the average MSE of $1.42 \times 10^{-4}$ for LDPRecover versus $8.78 \times 10^{-2}$ for the poisoned frequencies.

Overall, these observations illustrate the effectiveness and applicability of our proposed recovery methods. 

\section{Discussion}\label{Sec_discussion}
\subsection{Applicability to Other Aggregation Functions}
LDPRecover is designed for frequency estimation in LDP protocols. Its effectiveness is based on the principle that frequencies aggregated by LDP protocols are normally distributed when the number of users is sufficiently large, as supported by Theorem 1. This foundational principle ensures that if other aggregation functions (e.g., mean estimation) can be decomposed to frequency estimation problems, LDPRecover retains its effectiveness. As a case in point, consider Harmony~\cite{nguyen2016collecting}, a common LDP protocol for mean estimation. Harmony discretizes numerical values into binary categories (e.g., $1$ or $-1$) and applies Randomized Response (an LDP protocol for frequency) for perturbation, subsequently aggregating these perturbed frequencies to compute the mean. Since Harmony follows the frequency estimation paradigm, LDPRecover is effective for Harmony.

\subsection{Extension to Defend against Input Poisoning Attacks}
\begin{figure}[tb]
	\centerline{\includegraphics[width=\linewidth]{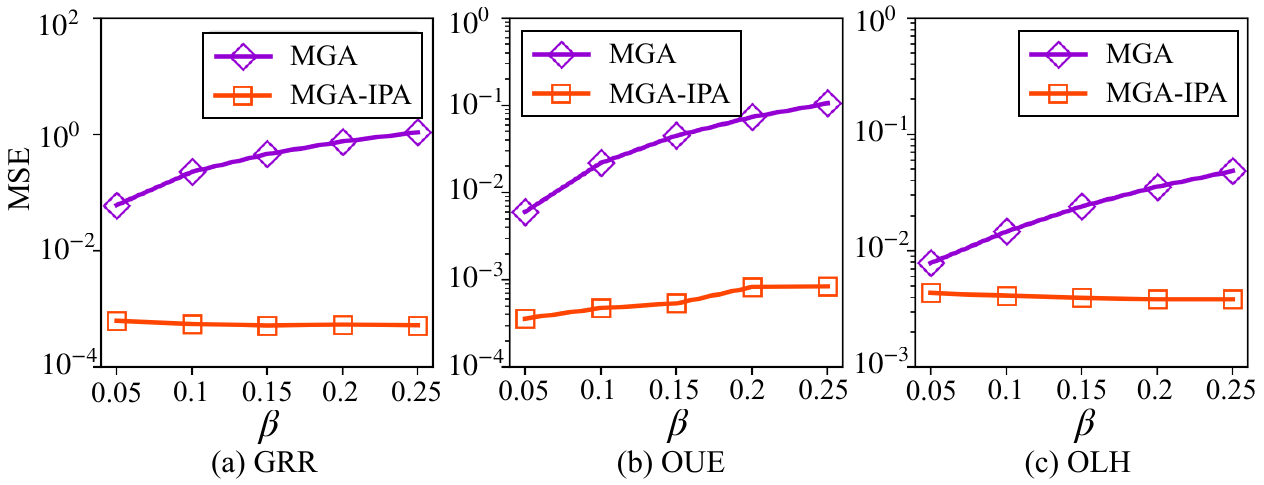}}
	\caption{Comparison of MGA performance under the general poisoning attack and IPA (IPUMS).}
	\label{fig_ipa_opa_beta}
\end{figure}

Note that our threat model essentially follows the general poisoning attack~\cite{cao2021data,cheu2021manipulation,li2023fine, du2023differential}, where malicious users can send attacker-crafted data directly to the server, bypassing LDP perturbation mechanisms. Some works~\cite{cao2021data,cheu2021manipulation,li2023fine, du2023differential} have also explored the input poisoning attack (IPA) you mentioned, where malicious users strictly follow the LDP perturbation. While IPA makes frequency recovery more challenging due to perturbation, it is far less effective than general poisoning attacks, as highlighted by the majority of existing studies~\cite{cao2021data,cheu2021manipulation,li2023fine, du2023differential}. To verify this, we implemented MGA under IPA (denoted as MGA-IPA) by sending the MGA-produced malicious data to the server after perturbing it via LDP perturbation mechanisms, and then compared its performance difference with the original MGA (under the general poisoning attack). The experimental results in Figure \ref{fig_ipa_opa_beta} show that MGA-IPA performs significantly worse than the original MGA (under the general poisoning attack). For example, when attacking GRR, the MSE of original MGA is between $6.07 \times 10^{-2} \sim 1.08$ while that of MGA-IPA alone is between $5.16 \times 10^{-4} \sim 6.21 \times 10^{-4}$, resulting in an improvement of $2 \sim 4$ orders of magnitude.

\begin{figure}[tb]
	\setlength{\abovecaptionskip}{-0.5em}
	\setlength{\belowcaptionskip}{-0.5em}
	\centerline{\includegraphics[width=\linewidth]{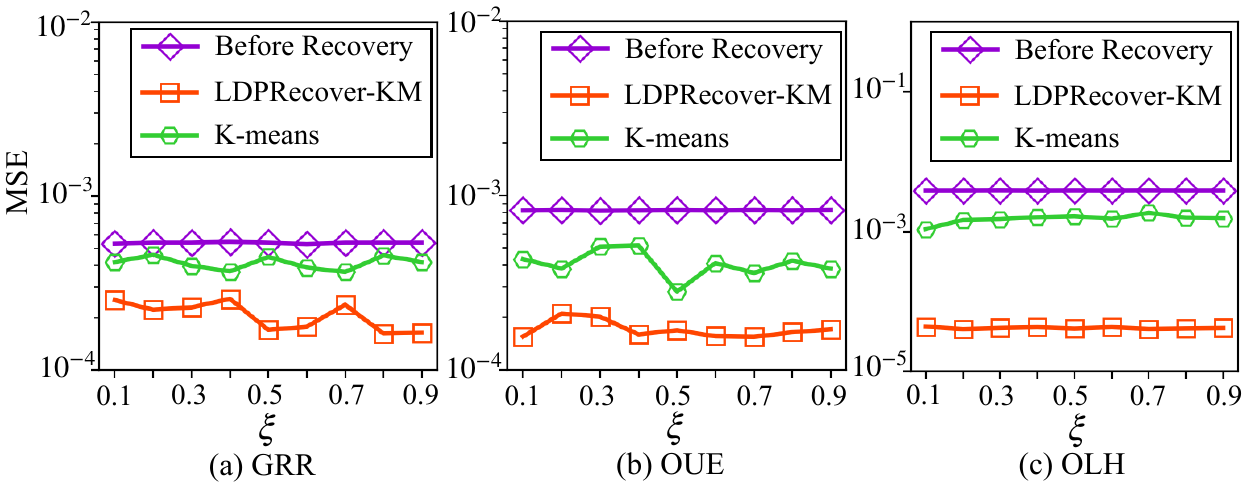}}
	\caption{Comparison of LDPRecover-KM and k-means performance under MGA-IPA (IPUMS).}
	\label{fig_rec_ipa_sr}
\end{figure}

Although our primary focus is on defending against general poisoning attacks, LDPRecover can be adapted to counteract IPA by integrating existing detection methods, such as k-means clustering approach \cite{li2023fine,du2023differential}. The k-means-based defense samples multiple subsets from users and clusters these subsets into two clusters: the cluster with more subsets is considered as the genuine cluster and are used to frequency estimation, while the other cluster is considered as the malicious cluster. Note that under IPA, we cannot estimate the statistics of malicious frequencies in the same way as in this work (e.g., Equation (\ref{equ_estimate_sum_fy})), since the statistics of malicious data align with the genuine data. To address this problem, we integrate k-means-based defense into LDPRecover (denoted by LDPRecover-KM): we use k-means-based defense to estimate the cluster of malicious users and learn statistics of malicious frequencies from the cluster, making the proposed LDPRecover still effective. The results are shown in Figure \ref{fig_rec_ipa_sr}, where $\xi$ is the sample rate of k-means-based defense. This figure shows that by such integration, LDPRecover can recover accurate aggregated frequencies under IPA. For example in Figure \ref{fig_rec_ipa_sr} (a), when MGA-IPA attacks GRR, the integration of LDPRecover with k-means clustering yields a 48.9\% improvement in recovery accuracy compared to using k-means clustering alone.

\subsection{Applicability to Multi-Attacker Case}
\begin{figure}[tb]
	\setlength{\abovecaptionskip}{-0.5em}
	\setlength{\belowcaptionskip}{-0.5em}
	\centerline{\includegraphics[width=\linewidth]{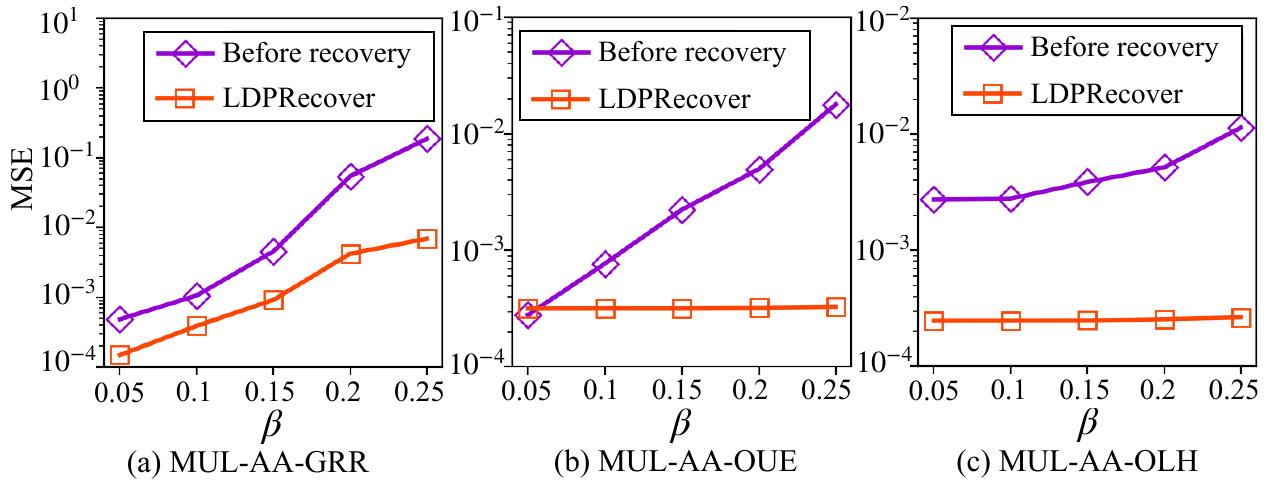}}
	\caption{The performance of LDPRecover when dealing with multi-attacker poisoning attacks (IPUMS).}
	\label{fig_mul_att_beta}
\end{figure}

As a final note, LDPRecover can also be used to defend against the multi-attacker threat model, in which multiple attackers control different groups of malicious users. Specifically, under the adaptive attack in LDPRecover, multiple attackers sampling malicious data from their respective attacker-designed distributions can be viewed as one attacker sampling malicious data from the joint distribution of these distributions. That is, the multi-attacker threat model can be treated as a case of the original threat model (in Section IV.A), so LDPRecover is still effective for this case. To verify this, we conduct the following experiments to evaluate the performance of LDPRecover when dealing with multiple attackers. In the experiment, we set up five attackers to perform AA and randomly assign malicious users to these attackers. Figure \ref{fig_mul_att_beta} demonstrates that LDPRecover accurately recovers aggregated frequencies from multi-attacker poisoned data. For example in Figure \ref{fig_mul_att_beta} (a), LDPRecover achieves an average improvement of 80.2\% in the accuracy of aggregated frequencies compared to the poisoned data. These results validate our analysis above.

\section{Conclusion}\label{Sec_conclusion}
In this work, we perform the first systematic study on frequency recovery from LDP poisoning attacks. Our proposed recovery method, called LDPRecover, can eliminate the impact of poisoning attacks on the aggregated frequencies gathered from LDP protocols. In particular, LDPRecover can be used as a frequency recovery paradigm that can enhance the overall accuracy of the recovered frequencies by integrating the attack details into LDPRecover. Our experimental results confirm the effectiveness of the proposed method at recovering aggregated frequencies from the poisoned ones. 

An interesting future work is to extend LDPRecover to poisoning attacks on LDP protocols for more complex tasks, such as key-value pairs collection under LDP. 

\section*{Acknowledgment}
This work was supported by the National Key Research and Development Program of China (No. 2022ZD0120200), the National Natural Science Foundation of China (Grant No: 62102334, 92270123 and 62372122), and the Research Grants Council, Hong Kong SAR, China (Grant No: 15225921, 15209922, 15208923, 15210023 and C2004-21GF), the Fundamental Research Funds for the Central Universities (NO. 501QYJC2023121001).

\bibliographystyle{IEEEtran}
\bibliography{IEEEabrv,references}

\begin{thebibliography}{10}
\providecommand{\url}[1]{#1}
\csname url@samestyle\endcsname
\providecommand{\newblock}{\relax}
\providecommand{\bibinfo}[2]{#2}
\providecommand{\BIBentrySTDinterwordspacing}{\spaceskip=0pt\relax}
\providecommand{\BIBentryALTinterwordstretchfactor}{4}
\providecommand{\BIBentryALTinterwordspacing}{\spaceskip=\fontdimen2\font plus
\BIBentryALTinterwordstretchfactor\fontdimen3\font minus
  \fontdimen4\font\relax}
\providecommand{\BIBforeignlanguage}[2]{{%
\expandafter\ifx\csname l@#1\endcsname\relax
\typeout{** WARNING: IEEEtran.bst: No hyphenation pattern has been}%
\typeout{** loaded for the language `#1'. Using the pattern for}%
\typeout{** the default language instead.}%
\else
\language=\csname l@#1\endcsname
\fi
#2}}
\providecommand{\BIBdecl}{\relax}
\BIBdecl

\bibitem{duchi2013local}
J.~C. Duchi, M.~I. Jordan, and M.~J. Wainwright, ``Local privacy and
  statistical minimax rates,'' in \emph{FOCS}, 2013.

\bibitem{dwork2006calibrating}
C.~Dwork, F.~McSherry, K.~Nissim, and A.~Smith, ``Calibrating noise to
  sensitivity in private data analysis,'' in \emph{TCC}, 2006.

\bibitem{dwork2014algorithmic}
C.~Dwork, A.~Roth \emph{et~al.}, ``The algorithmic foundations of differential
  privacy,'' \emph{Foundations and Trends in Theoretical Computer Science},
  2014.

\bibitem{erlingsson2014rappor}
{\'U}.~Erlingsson, V.~Pihur, and A.~Korolova, ``Rappor: Randomized aggregatable
  privacy-preserving ordinal response,'' in \emph{CCS}, 2014.

\bibitem{fanti2015building}
G.~Fanti, V.~Pihur, and {\'U}.~Erlingsson, ``Building a rappor with the
  unknown: Privacy-preserving learning of associations and data dictionaries,''
  2016.

\bibitem{bittau2017prochlo}
A.~Bittau, {\'U}.~Erlingsson, P.~Maniatis, I.~Mironov, A.~Raghunathan, D.~Lie,
  M.~Rudominer, U.~Kode, J.~Tinnes, and B.~Seefeld, ``Prochlo: Strong privacy
  for analytics in the crowd,'' in \emph{SOSP}, 2017.

\bibitem{appledf2017}
A.~D.~P. Team, ``Learning with privacy at scale,'' \emph{Machine Learning
  Journal}, 2017.

\bibitem{cheu2021manipulation}
A.~Cheu, A.~Smith, and J.~Ullman, ``Manipulation attacks in local differential
  privacy,'' in \emph{S\&P}, 2021.

\bibitem{cao2021data}
X.~Cao, J.~Jia, and N.~Z. Gong, ``Data poisoning attacks to local differential
  privacy protocols,'' in \emph{USENIX Security}, 2021.

\bibitem{wu2022poisoning}
Y.~Wu, X.~Cao, J.~Jia, and N.~Z. Gong, ``Poisoning attacks to local
  differential privacy protocols for key-value data,'' 2022.

\bibitem{kieu2018outlier}
T.~Kieu, B.~Yang, and C.~S. Jensen, ``Outlier detection for multidimensional
  time series using deep neural networks,'' in \emph{MDM}, 2018.

\bibitem{zhou2018non}
Y.~Zhou, H.~Zou, R.~Arghandeh, W.~Gu, and C.~J. Spanos, ``Non-parametric
  outliers detection in multiple time series a case study: Power grid data
  analysis,'' in \emph{AAAI}, 2018.

\bibitem{su2019robust}
Y.~Su, Y.~Zhao, C.~Niu, R.~Liu, W.~Sun, and D.~Pei, ``Robust anomaly detection
  for multivariate time series through stochastic recurrent neural network,''
  in \emph{KDD}, 2019.

\bibitem{kairouz2014extremal}
P.~Kairouz, S.~Oh, and P.~Viswanath, ``Extremal mechanisms for local
  differential privacy,'' in \emph{NeurIPS}, 2014.

\bibitem{wang2017locally}
T.~Wang, J.~Blocki, N.~Li, and S.~Jha, ``Locally differentially private
  protocols for frequency estimation,'' in \emph{USENIX Security}, 2017.

\bibitem{warner1965randomized}
S.~L. Warner, ``Randomized response: A survey technique for eliminating evasive
  answer bias,'' \emph{Journal of the American Statistical Association}, 1965.

\bibitem{bassily2015local}
R.~Bassily and A.~Smith, ``Local, private, efficient protocols for succinct
  histograms,'' in \emph{STOC}, 2015.

\bibitem{kairouz2016discrete}
P.~Kairouz, K.~Bonawitz, and D.~Ramage, ``Discrete distribution estimation
  under local privacy,'' in \emph{International Conference on Machine
  Learning}.\hskip 1em plus 0.5em minus 0.4em\relax PMLR, 2016, pp. 2436--2444.

\bibitem{zhang2018calm}
Z.~Zhang, T.~Wang, N.~Li, S.~He, and J.~Chen, ``Calm: Consistent adaptive local
  marginal for marginal release under local differential privacy,'' in
  \emph{CCS}, 2018.

\bibitem{jia2019calibrate}
J.~Jia and N.~Z. Gong, ``Calibrate: Frequency estimation and heavy hitter
  identification with local differential privacy via incorporating prior
  knowledge,'' in \emph{INFOCOM}, 2019.

\bibitem{wang2019consistent}
T.~Wang, M.~Lopuha{\"a}-Zwakenberg, Z.~Li, B.~Skoric, and N.~Li, ``Locally
  differentially private frequency estimation with consistency,'' in
  \emph{NDSS}, 2020.

\bibitem{xue2022ddrm}
Q.~Xue, Q.~Ye, H.~Hu, Y.~Zhu, and J.~Wang, ``Ddrm: A continual frequency
  estimation mechanism with local differential privacy,'' \emph{IEEE
  Transactions on Knowledge and Data Engineering}, 2022.

\bibitem{ding2017collecting}
B.~Ding, J.~Kulkarni, and S.~Yekhanin, ``Collecting telemetry data privately,''
  in \emph{NeurIPS}, 2017.

\bibitem{duchi2018minimax}
J.~C. Duchi, M.~I. Jordan, and M.~J. Wainwright, ``Minimax optimal procedures
  for locally private estimation,'' \emph{Journal of the American Statistical
  Association}, 2018.

\bibitem{wang2019collecting}
N.~Wang, X.~Xiao, Y.~Yang, J.~Zhao, S.~C. Hui, H.~Shin, J.~Shin, and G.~Yu,
  ``Collecting and analyzing multidimensional data with local differential
  privacy,'' in \emph{ICDE}, 2019.

\bibitem{li2020estimating}
Z.~Li, T.~Wang, M.~Lopuha{\"a}-Zwakenberg, N.~Li, and B.~{\v{S}}koric,
  ``Estimating numerical distributions under local differential privacy,'' in
  \emph{SIGMOD}, 2020.

\bibitem{duan2022utility}
J.~Duan, Q.~Ye, and H.~Hu, ``Utility analysis and enhancement of ldp mechanisms
  in high-dimensional space,'' 2022.

\bibitem{bassily2017practical}
R.~Bassily, K.~Nissim, U.~Stemmer, and A.~G. Thakurta, ``Practical locally
  private heavy hitters,'' in \emph{NeurIPS}, 2017.

\bibitem{wang2019locally}
T.~Wang, N.~Li, and S.~Jha, ``Locally differentially private heavy hitter
  identification,'' \emph{TDSC}, 2019.

\bibitem{qin2016heavy}
Z.~Qin, Y.~Yang, T.~Yu, I.~Khalil, X.~Xiao, and K.~Ren, ``Heavy hitter
  estimation over set-valued data with local differential privacy,'' in
  \emph{CCS}, 2016.

\bibitem{wang2018locally}
T.~Wang, N.~Li, and S.~Jha, ``Locally differentially private frequent itemset
  mining,'' in \emph{S\&P}, 2018.

\bibitem{li2014data}
C.~Li, M.~Hay, G.~Miklau, and Y.~Wang, ``A data-and workload-aware algorithm
  for range queries under differential privacy,'' \emph{PVLDB}, 2014.

\bibitem{cormode2019answering}
G.~Cormode, T.~Kulkarni, and D.~Srivastava, ``Answering range queries under
  local differential privacy,'' \emph{PVLDB}, 2019.

\bibitem{yang2020answering}
J.~Yang, T.~Wang, N.~Li, X.~Cheng, and S.~Su, ``Answering multi-dimensional
  range queries under local differential privacy,'' \emph{PVLDB}, 2020.

\bibitem{he2015dpt}
X.~He, G.~Cormode, A.~Machanavajjhala, C.~M. Procopiuc, and D.~Srivastava,
  ``Dpt: differentially private trajectory synthesis using hierarchical
  reference systems,'' \emph{PVLDB}, 2015.

\bibitem{chen2016private}
R.~Chen, H.~Li, A.~K. Qin, S.~P. Kasiviswanathan, and H.~Jin, ``Private spatial
  data aggregation in the local setting,'' in \emph{ICDE}, 2016.

\bibitem{qin2017generating}
Z.~Qin, T.~Yu, Y.~Yang, I.~Khalil, X.~Xiao, and K.~Ren, ``Generating synthetic
  decentralized social graphs with local differential privacy,'' in \emph{CCS},
  2017.

\bibitem{avent2017blender}
B.~Avent, A.~Korolova, D.~Zeber, T.~Hovden, and B.~Livshits, ``{BLENDER}:
  Enabling local search with a hybrid differential privacy model,'' in
  \emph{USENIX Security}, 2017.

\bibitem{cormode2018marginal}
G.~Cormode, T.~Kulkarni, and D.~Srivastava, ``Marginal release under local
  differential privacy,'' in \emph{SIGMOD}, 2018.

\bibitem{ren2018textsf}
X.~Ren, C.-M. Yu, W.~Yu, S.~Yang, X.~Yang, J.~A. McCann, and S.~Y. Philip,
  ``{LoPub}: High-dimensional crowdsourced data publication with local
  differential privacy,'' \emph{TIFS}, 2018.

\bibitem{gursoy2018utility}
M.~E. Gursoy, L.~Liu, S.~Truex, L.~Yu, and W.~Wei, ``Utility-aware synthesis of
  differentially private and attack-resilient location traces,'' in \emph{CCS},
  2018.

\bibitem{wang2019answering}
T.~Wang, B.~Ding, J.~Zhou, C.~Hong, Z.~Huang, N.~Li, and S.~Jha, ``Answering
  multi-dimensional analytical queries under local differential privacy,'' in
  \emph{SIGMOD}, 2019.

\bibitem{ye2020towards}
Q.~Ye, H.~Hu, M.~H. Au, X.~Meng, and X.~Xiao, ``Towards locally differentially
  private generic graph metric estimation,'' in \emph{CDE}, 2020.

\bibitem{ye2020lf}
Q.~\vspace{0mm}Ye, H.~Hu, M.~H. Au, X.~Meng, and X.~Xiao, ``Lf-gdpr: A
  framework for estimating graph metrics with local differential privacy,''
  \emph{TKDE}, 2020.

\bibitem{cunningham2021real}
T.~Cunningham, G.~Cormode, H.~Ferhatosmanoglu, and D.~Srivastava, ``Real-world
  trajectory sharing with local differential privacy,'' \emph{PVLDB}, 2021.

\bibitem{shanthikumar1984central}
J.~Shanthikumar and U.~Sumita, ``A central limit theorem for random sums of
  random variables,'' \emph{Operations Research Letters}, 1984.

\bibitem{fischer2011history}
H.~Fischer, \emph{A history of the central limit theorem: from classical to
  modern probability theory}.\hskip 1em plus 0.5em minus 0.4em\relax Springer.

\bibitem{sunxy}
\BIBentryALTinterwordspacing
X.~Sun, Q.~Ye, H.~Hu, J.~Duan, T.~Wo, J.~Xu, and R.~Yang, ``Ldprecover:
  Recovering frequencies from poisoning attacks against local differential
  privacy,'' Tech. Rep. [Online]. Available:
  \url{https://www.dropbox.com/scl/fo/88to91xn1u368u6dxoyje/h?rlkey=ttp8zopq0htbkrcbdyeuieqwm&dl=0}
\BIBentrySTDinterwordspacing

\bibitem{chen2021private}
R.~Chen, H.~Li, S.~Kasiviswanathan, and H.~Jin, ``Private dataaggregation
  framework for untrusted servers,'' Mar.~23 2021, uS Patent 10,956,603.

\bibitem{hay2009boosting}
M.~Hay, V.~Rastogi, G.~Miklau, and D.~Suciu, ``Boosting the accuracy of
  differentially-private histograms through consistency,'' \emph{PVLDB}, 2010.

\bibitem{karush1939minima}
W.~Karush, ``Minima of functions of several variables with inequalities as side
  constraints,'' \emph{M. Sc. Dissertation. Dept. of Mathematics, Univ. of
  Chicago}, 1939.

\bibitem{kuhn2014nonlinear}
H.~W. Kuhn and A.~W. Tucker, ``Nonlinear programming,'' in \emph{Traces and
  emergence of nonlinear programming}, 2014.

\bibitem{ipums}
S.~Ruggles, S.~Flood, R.~Goeken, M.~Schouweiler, and M.~Sobek, ``Ipums usa:
  Version 12.0 [dataset]. minneapolis, mn: Ipums, 2022,''
  \emph{https://doi.org/10.18128/D010.V12.0}, 2022.

\bibitem{fire}
``San francisco fire department calls for service,''
  \emph{http://bit.ly/336sddL}, 2023.

\bibitem{nguyen2016collecting}
T.~T. Nguy{\^e}n, X.~Xiao, Y.~Yang, S.~C. Hui, H.~Shin, and J.~Shin,
  ``Collecting and analyzing data from smart device users with local
  differential privacy,'' \emph{arXiv preprint arXiv:1606.05053}, 2016.

\bibitem{li2023fine}
X.~Li, N.~Li, W.~Sun, N.~Z. Gong, and H.~Li, ``Fine-grained poisoning attack to
  local differential privacy protocols for mean and variance estimation,'' in
  \emph{USENIX Security'23}, 2023, pp. 1739--1756.

\bibitem{du2023differential}
R.~Du, Q.~Ye, Y.~Fu, H.~Hu, J.~Li, C.~Fang, and J.~Shi, ``Differential
  aggregation against general colluding attackers,'' in \emph{ICDE}, 2023.

\end{thebibliography}

\end{document}